\documentclass[aps,prl,reprint,superscriptaddress,nofootinbib]{revtex4-2}
\pdfoutput=1

\usepackage{booktabs}
\usepackage[english]{babel}
\usepackage{amsmath,amssymb,amsbsy,amstext, amsthm, simplewick}
\usepackage{hyperref}
\usepackage{graphicx}
\usepackage{amsfonts}
\usepackage{amssymb}
\usepackage{upgreek}
\usepackage[svgnames,dvipsnames,x11names,table]{xcolor}
\usepackage{multirow}
\usepackage{bm}
\usepackage{braket}
\usepackage{mathtools}
\usepackage{cleveref}
\usepackage{comment}
\usepackage{scalerel}
\usepackage[normalem]{ulem}
\usepackage{slashed}
\usepackage{enumitem}
\usepackage{dsfont}
\usepackage{tikz}
\usetikzlibrary{decorations.markings}
\usetikzlibrary{shapes.misc}

\makeatletter
\g@addto@macro\bfseries{\boldmath}
\makeatother

\hypersetup{
    colorlinks=true,
    linkcolor={red!50!black},
    citecolor={blue!50!black},
    urlcolor={blue!80!black}
}

\usepackage{colortbl}

\makeatletter
\newlength{\apb@width}
\newcommand{\autoparbox}[2][c]{\settowidth{\apb@width}{#2}\parbox[#1]{\apb@width}{#2}}

\makeatother
\definecolor{lightgray}{gray}{0.9}

\usepackage[framemethod=default]{mdframed}
\newmdenv[skipabove=7pt,
skipbelow=7pt,
rightline=false,
leftline=false,
topline=false,
bottomline=false,
backgroundcolor=gray!10,
linecolor=gray,
innerleftmargin=5pt,
innerrightmargin=5pt,
innertopmargin=5pt,
innerbottommargin=5pt,
leftmargin=0cm,
rightmargin=0cm,
linewidth=4pt]{eBox}

\usepackage[most]{tcolorbox}
\tcbset{colback=white, colframe=black,
        highlight math style= {enhanced, 
            colframe=red,colback=red!10!white,boxsep=0pt}
        }
\definecolor{light-gray}{gray}{0.95}

\crefname{table}{Table}{Tables}
\crefname{equation}{Eq.}{Eqs.}
\crefname{appendix}{App.}{Apps.}
\crefname{section}{Sec.}{Secs.}
\crefname{figure}{Fig.}{Figs.}

\def\beq{\begin{equation}}
\def\eeq{\end{equation}}

\def\bea{\begin{eqnarray}}
\def\eea{\end{eqnarray}}

\def\beq{\begin{equation}}
\def\eeq{\end{equation}}
\def\bea{\begin{eqnarray}}
\def\eea{\end{eqnarray}}

\def\O{{\cal O}}

\def\P{{{\vec P}}}

\def\k{{\vec{k}}}

\def\p{{\vec p}}

\def\x{{\vec x}}

\def\X{{\vec X}}

\DeclareRobustCommand{\SkipTocEntry}[4]{}

\definecolor{colorTC}{rgb}{.2,.7,.2}

\definecolor{amethyst}{rgb}{0.6, 0.4, 0.8}

\definecolor{acolor}{rgb}{0.4, 0.2, 0.4}

\definecolor{blue3}{RGB}{31, 119, 180}
\definecolor{red3}{RGB}{	214, 39, 40}
\definecolor{orange3}{RGB}{255, 127, 14}
\definecolor{green3}{RGB}{44, 160, 44}

\newcommand{\prlsection}[1]{\vspace{2pt}\noindent\textit{#1.}---}

\begin{document}

\title{Nambu-Goldstone Bosons and Quantum Fisher Information}

\author{Eugene Chen}
\affiliation{Department of Physics, University of California at San Diego,  La Jolla, CA 92093, USA}
\author{Daniel Green}
\affiliation{Department of Physics, University of California at San Diego,  La Jolla, CA 92093, USA}

\begin{abstract}
The quantum Fisher matrix and geometric tensor are powerful tools for quantifying the nature of quantum systems. Unfortunately, the formalism needed to compute these quantities is opaque and has limited the development of physical intuition for their behavior, even in the ground state. For quantum systems with continuous symmetries, we know the ground state can be understood in a model-independent and intuitive way from the correlation functions of conserved currents and charges. In this paper, we demonstrate an equivalence between the quantum Fisher matrix associated with the ground states of systems with continuous symmetries and the Nambu-Goldstone bosons produced by broken generators. We will show that the information metric defined by the Fisher matrix is the same as the Zamolodchikov metric that defines the coset space of the system. Next we show that the Uhlmann curvature associated with multi-parameter estimation is the same as the charge density matrix that defines the difference between type-A and type-B Nambu-Goldstone bosons. We demonstrate these connections in several examples and discuss the generalization to finite temperature.
\end{abstract}

\maketitle

\prlsection{Introduction} 
The nature of the ground state of large quantum systems is an area of great importance across physics~\cite{Zeng:2015pxf}. In recent years, there has been significant progress using tools from quantum information theory to characterize these states via their entanglement entropies and related quantities~\cite{Laflorencie:2015eck,Camasca:2026awd}. Of particular note has been the use of strong sub-additivity of entanglement entropies in proving monotonicity of central charges under renormalization group (RG) flow~\cite{Casini:2022rlv}. Moreover, it has become the focus of our understanding of quantum gravity, particularly in black hole and cosmological backgrounds~\cite{Bousso:2022ntt}.

The quantum Fisher information (QFI)~\cite{PhysRevLett.72.3439,Paris:2008zgg}, or more generally the quantum geometric tensor~\cite{Provost:1980nc,Kolodrubetz:2017ofs}, is another information-theoretic quantity that is capable of characterizing properties of these states.  It is broadly useful defining the information geometry of pure states and can be used to characterize the physical properties of the vacuum. The QFI has the added benefit of being directly tied to more conventional observables, such as the means and variances of local operators. From the Holevo-Cram\'er-Rao bound~\cite{Holevo:2011}, the quantum geometric tensor defines the limiting ability of experimentalists to probe the system.

As a practical tool, the QFI formalism can be counter-intuitive. Given a density matrix, $\rho$, and a set of parameters $\theta_a$, the quantum Fisher matrix (QFM), $F_{ab}$, is defined in terms of the symmetric logarithmic derivatives (SLD), $ L_a$, as 
\beq\label{eq:QFM}
\frac{\partial}{\partial \theta_a} \rho =\frac{1}{2} \{  L_a , \rho \} \to F_{ab} = \frac{1}{2} {\rm Tr} \rho \{ L_a,  L_b \} \ .
\eeq 
Although $F_{ab}$ defines the quantum Cram\'er-Rao bound, it can only be saturated for multi-parameter estimation when ${\rm Tr}\rho[L_a,L_b]=0$~\cite{Ragy:2016qji}. For commuting observables, we can work in the eigenbasis of the $\theta_a$ parameters such that the Fisher matrix is purely classical. In this sense, the QFI formalism is more difficult to use than its classical counter-part, and yet it seems to be most useful in the classical limit. Like the CHSH inequalities~\cite{Clauser:1969ny}, truly quantum aspects of measurement appear when ${\rm Tr} \rho [ L_a,  L_b] \neq 0$~\cite{Valenti:2019znw} (via the imaginary parts of the quantum geometric tensor) and show up in the Holevo-Cram\'er-Rao  bound for multi-parameter non-commuting measurements.

We can gain a lot of useful intuition about the relationship between the QFI and the nature of states in systems with continuous symmetries. Suppose that the parameters $\theta_a$ can be shifted by acting on the density matrix with conserved charges $ Q_a$, so that
\beq\label{eq:rhoQ}
\rho_\theta = e^{-i \sum_a \theta_a  Q_a} \rho_0 e^{i \sum_a \theta_a  Q_a} \ .
\eeq
As a concrete example, we might be trying to measure the position of a particle, $x$, where the location can be shifted by acting with the conserved momentum, $ p$, via $\rho(x-\Delta x, x'-\Delta x) =\langle x| e^{-i \Delta x  p} \rho e^{i \Delta x  p}|x'\rangle$.
Notice, however, that if we put the system in a pure state that is annihilated by the charge, $ Q_a |\psi\rangle =0$, then $ L_a = 0$ and we will find there is no Fisher information associated with $\theta_a$. In this precise sense, we will see that the QFI reflects the breaking of the symmetry $ Q_a$ in the state $|\psi\rangle$~\cite{Marvian:2014tjm,Kudo:2022jan,Wilson:2025kjd}. In practice, we don't measure $\theta_a$ defined over an infinite volume and therefore we need to extend this intuition to measurements over finite regions or with finite momenta.

In quantum field theory (QFT), this connection between Fisher information and spontaneous symmetry breaking further implies a connection to the theory of Nambu-Goldstone bosons (NGBs). NGBs~\cite{Nambu:1961tp,Goldstone:1961eq,Goldstone:1962es} are gapless degrees of freedom that must arise when continuous global symmetries are spontaneously broken. NGBs are created when the charges, or associated currents, act on the vacuum. We have already seen that the Fisher information is similarly related to these charges. As such, it is natural to guess that there is a general relationship between the current correlators that define the NGBs and those that define QFI. Our goal here is to construct this map between Goldstone modes and the quantum geometric tensor. Moreover, we will connect the Holevo-Cram\'er-Rao bound directly to the pattern of symmetry breaking in both relativistic and non-relativistic theories. In particular, we will see that the imaginary part of the quantum geometric tensor, or the Uhlmann curvature~\cite{UHLMANN1986229}, is precisely the charge density matrix that distinguishes type-A and type-B NGBs~\cite{Schafer:2001bq,Nambu:2004yia,Watanabe:2011ec,Hidaka:2012ym,Watanabe:2012hr,Watanabe:2014fva}. We will then show that the geometric interpretation of the QFI is identical to the coset space.

One of the potential applications of this work is the use of QFI as a probe of symmetry breaking~\cite{Zanardi:2007mqd,Venuti:2007izd,Gu:2008dyh,Kolodrubetz:2013mla,Ares:2022koq,Yamashika:2025ghp,Yamaguchi:2026mqp,simeth2026}. Given that the Fisher information is a quantity one can calculate in any state, it has been proposed as a simple diagnostic of phase transitions. The connection between NGBs and the QFI enables a more precise accounting of how we are testing the state of the system through quantum measurement. 
Beyond this practical application, our broader ambition is to better understand how quantum mechanics manifests itself in the fundamental limits of measurement in complex systems. Multi-parameter estimation of non-commuting observables is at the heart of quantum measurement, and yet lies beyond the reach of the QFI alone.

This paper is organized as follows. First we show how symmetry breaking manifests itself in both the theory of Nambu-Goldstone bosons and the quantum geometric tensor. We will then match the two descriptions including the information and Zamolodchikov metrics. We will then demonstrate how this map works in the $O(N)$ and $SO(3)\times U(1)$ models, illustrating type-A and type-B NGBs, respectively. We will finally discuss generalizations to mixed states before concluding.

\prlsection{Two Views of Symmetry Breaking}
We wish to understand the nature of the ground state of a $(d+1)$-dimensional quantum system $|\Omega\rangle$ with a group of continuous global symmetries ${\cal G}$ and associated conserved currents $j_A^{\mu}$, or charges $Q_A = \int d^d x j^0_A(\x)$, for each generator $T_A$. One way to define spontaneous symmetry breaking is to assume there is an operator $\phi_\ell(\x)$, where $\ell=1,.., M$ is the index associated with the representation under ${\cal G}$, that acquires a vacuum expectation value
\beq
\langle \phi_\ell \rangle = v_\ell \to \langle [Q_a, \phi_\ell ]\rangle = -i \sum_{\ell'} [T_a]_{\ell \ell'} v_{\ell'}  \neq 0 \ ,
\eeq
where $a$ is the list of broken generators defined by the non-vanishing RHS. Clearly a non-zero result requires that broken charges act non-trivially on the vacuum,  $ Q_a |\Omega\rangle\neq 0$.

This setup is the precise context in which gapless NGBs appear. Suppose we act on the vacuum with a smeared current
\beq
 Q_{a}(\k) = \int_V d^d x e^{i\k\cdot \x} j_a^0(\x) \to |a,\k \rangle =  Q_{a}(\k) |\Omega\rangle \ ,
\eeq
where $V$ will define an effectively finite volume of integration. Since the current is conserved, the energy of this state is 
\beq
  [H ,  Q_{a}(\k) ]|\Omega \rangle =  \k\cdot \vec j_a(\k) |\Omega \rangle  \approx E(\k)  Q_a(\k)|\Omega\rangle\ ,
\eeq
where $\vec j(\k)=\int d^d x \vec j_a(\x) e^{i\k\cdot \x}$. Here the notion of energy $E(k)$ is only approximate because $ Q_a(\k) |\Omega\rangle$ is not an energy eigenstate in general. Taking $V \to \infty$ and $\k \to 0$ then implies $E(\k) \to 0$. In this sense, spontaneously broken symmetries generate gapless states. For relativistic theories, one can show that this state always contains a single-particle state obeying $E(k)^2= k^2$.

In both relativistic and non-relativistic theories, one defines a field associated with these Goldstone modes, $\pi^a(\x,t)$. The definition is ambiguous in general, but a common choice is for them to transform non-linearly so that $[Q_a, \pi^a(\x,t) ] = -i + {\cal O}(\pi)$. With this definition, we see that the charge contains the conjugate momentum, $Q_a \supset \Pi_a$. In this way, the NGBs define a coordinate space on the space of degenerate vacua created by the charges. Much about the physics of spontaneous symmetry breaking is defined by the geometry of this space. At the same time, to gain a model-independent understanding of spontaneous symmetry breaking, it is useful to work with the currents and charges, rather than $\pi_a$, because their definition is unambiguous (up to ``improvement terms"). Defining an order parameter $X_a = \vec\phi^T T^a \vec v$, and matrix elements with the NGB state $|\pi_b,\k\rangle$, 
\beq
\Phi_{ab}(\k) = \langle X_a |\pi_b,\k\rangle \qquad {\cal J}_{ab}(\k) = \langle j^0_a |\pi_b,\k\rangle \ ,
\eeq
the two-point correlations involving the charges are 
\begin{align}
-i \kappa_{ab} \equiv \langle [Q_a(\k), X_b(0) ] \rangle &= -i \vec v^T (T^a)^T T^b \vec v\\
&=2i{\rm Im} {\cal J}_{ac} \Phi^\dagger_b{}^{c}\\
\langle Q_a(\k) Q_b(-\k) \rangle &= V {\cal J}_{ac} {\cal J}^\dagger{}^c_b \ .
\end{align}
Notice that $\kappa_{ab}$ is symmetric and positive by construction.

The last non-trivial feature of NGBs is the charge commutator. Specifically,  
\beq
2 {\rm Im}\langle Q_a Q_b \rangle=  V 2 {\rm Im}  {\cal J}_{ac} {\cal J}^\dagger{}^c_b= -i \langle [Q_a, Q_b ] \rangle  = V \varrho_{ab} \ .
\eeq
For relativistic theories, invariance under boosts for $Q_a= j_a^0(\k\to 0)$ combined with the locality of the commutator requires that $\varrho_{ab} =0$. This gives rise to the result that there is an independent degree of freedom (and single particle) for each broken generator. For a non-relativistic theory, we can have $\varrho_{ab} \neq 0$ and it signals that two of the Nambu-Goldstone bosons are actually conjugate momenta of each other. This can be seen directly in the form of the linear kinetic term
\beq
{\cal L} \supset -\frac{1}{2} \varrho_{ab} \pi^a \dot \pi^b \to Q_a \propto \Pi_a-K^0_a \supset \varrho_{ab} \pi^b \ .
\eeq
where $K_a^\mu$ is the total-derivative term defined by $\delta\mathcal{L} = \partial_\mu K^\mu$.
As a consequence, we see that $Q_a$ creates both $\pi_a$ and $\pi_b$ NGBs, $\langle Q_a |\pi_{a,b}\rangle \neq 0$. In this sense, $\varrho_{ab}$ defines the difference between Type-A and Type-B NGBs, and also the commutation relation of charges.

Now let's view the same system through the lens of the quantum Fisher matrix. We want to measure the parameters $\theta_a$ defined by the same smeared charges,
\beq
|\theta_a \rangle = e^{-i \tfrac{1}{2} \sum_a \theta_a (Q_a(\k)+ Q^{\dagger}_a(\k))} |\Omega\rangle \ .
\eeq
Hereafter, $Q_a$  without an explicit momentum argument denotes the Hermitian combination $Q_a\equiv\frac{Q_a(\vec{k})+Q_a(-\vec{k})}{2}$.
The QFM, $F_{ab}$ is defined using \cref{eq:QFM} in terms of the SLD operators, $ L_a$.
Since this is a pure state, $ L_a$ and $F_{ab}$ are straightforward to derive and are given by
\beq
 L_a = -i 2 [Q_a, \rho] \to F_{ab} = 2 \langle \{ \Delta Q_a , \Delta Q_b \} \rangle \ ,
\eeq
where $\Delta Q_a =\frac{1}{2} (Q_a(\k) -\langle Q_a(\k) \rangle)+ {\rm h.c.}$. To see the utility of the QFM, we define a set of unbiased estimators, $\hat \theta_a$, for the parameters $\theta_a$ such that 
\begin{align}
&{\rm Tr}\rho \tilde X_a = 0 \qquad \frac{1}{2}{\rm Tr} \rho \{ L_a, \tilde X_b \} =\delta_{ab} \\
&\to  i\langle [Q_a, \tilde X_b] \rangle = \delta_{ab} \ ,
\end{align}
where $\tilde X_a =\hat \theta_a -\theta_a$. The quantum Cram\'er-Rao (QCR) bound then states that
\begin{align}
&\Sigma_{ab} = \frac{1}{2}{\rm Tr}\rho \{\tilde X_a,\tilde X_b \}  \to {\rm Tr} W \Sigma \geq {\rm Tr} W F^{-1}
\end{align}
where $W$ is any positive weight matrix. 

The QCR bound for multi-parameter estimation can only be saturated when the Uhlmann curvature vanishes, $U_{ab}=\frac{1}{2i}{\rm Tr} \rho[L_a,L_b] =0$. When $U_{ab} \neq 0$, it indicates that the observables associated with $\theta_a$ and $\theta_b$ don't commute and therefore cannot be measured simultaneously\footnote{Technically speaking, the Uhlmann curvature can vanish for non-commuting observables, but these operators will behave as if they commute when we take many copies of the density matrix (i.e.~using the quantum central limit theorem~\cite{Cushen_Hudson_1971}).}. Instead, for a pure state, the stronger bound that can be saturated is the Holevo-Cram\'er-Rao bound~\cite{Holevo:2011,Yamagata:2013nin},
\beq\label{eq:HCR}
{\rm Tr} W \Sigma \geq  {\rm min}_{X} \left[ {\rm Tr} W {\rm Re} Z +|| W^{1/2} {\rm Im} Z W^{1/2} ||_1 \right] \ ,
\eeq
where $Z= {\rm Tr}\rho  X_a  X_b$ and $X_a$ is an unbiased estimator. It is often the case that the optimal estimator is $\tilde X_a = F^{-1}_a{}^{b} L_b$ in which case the bound becomes
\begin{align}
&{\rm min}_{X} \left[ {\rm Tr} W {\rm Re} Z +|| W^{1/2} {\rm Im} Z W^{1/2} ||_1 \right] \nonumber \\
&={\rm Tr} W F^{-1} + || \sqrt{W} F^{-1} U F^{-1} \sqrt{W} ||_1 \ . \label{eq:HCR2}
\end{align}
Note that for our pure state, the curvature is related to the charge density matrix,
\beq
U_{ab}= \frac{1}{2i} {\rm Tr}\rho[ L_a,   L_b] =\frac{2}{i} \langle [\Delta Q_a, \Delta Q_b] \rangle = V \varrho_{ab}  \ .
\eeq
In the presence of both Type-A and B NGBs, \cref{eq:HCR2} may be further lowered for measurements involving $\hat X_b \to \hat X_b + c \Delta Q_a$ when $Q_b$ is a type-B and $Q_a$ is type-A. Since $L_a =- 2i [Q_a, \rho]$, this change to $\hat X_b$ would be associated with the SLD of a complex parameter $\theta_a \to \theta_a + i \phi_a$~\cite{Matsumoto:1997hd}, and would have an interpretation as quasi-NGBs~\cite{Nitta:2014jta}.

When $U_{ab}\neq 0$, we cannot saturate the Cram\'er-Rao bound because $\theta_a$ and $\theta_b$ are conjugate momenta and cannot be measured simultaneously. This is the same reason for the $2 n_{\rm NGB} =   n_{\rm BG}$ counting for type B NGBs, where $n_{\rm NGB}$ and $n_{\rm BG}$ are the number of Goldstone modes and broken generators respectively.

\prlsection{Matching} At a basic level, we can see that the matrix elements that define the NGB states also characterize the quantum geometric tensor. Concretely, we can identify
\beq
F_{ab} = 2V{\rm Re}{\cal J}_{ac} {\cal J}^\dagger{}^c_b - 4\langle Q_a \rangle \langle Q_b \rangle \qquad U_{ab} =  V \varrho_{ab} \ .
\eeq
In addition, the Nambu-Goldstone boson operator $X_a \to \pi^a$ plays the same role as the unbiased estimator $\tilde X_a$ in the theory of quantum measurement.

At face value, it might seem that we are only matching a small number of two-point statistics. However, in both the theory of NGBs and information, the geometry defined by $\pi^a$ or $\theta^a$ is an important universal quantity. Specifically, we would like to see that the quantum Fisher information metric and the Zamolodchikov metric match.

To establish this connection, we begin with the NGB EFT, where the NGB fields $\pi^a$ are explicitly defined, and their corresponding Noether currents are given by 
\begin{equation}
    j^\mu_a= \frac{\partial S_{\rm eff}}{\partial(\partial_\mu\pi^b(x))}\xi^b_a(\pi) - K^{\mu}_a(x)
\end{equation}
where $S_{\rm eff}$ is the action, $\xi^b(\pi)$ is the Killing vector in the coset space, and $K^\mu_a$ accounts for the total derivative in the symmetry variation of the Lagrangian.

For pure states, we can express the NGB fields and their corresponding currents in terms of creation and annihilation operators, $\pi = Aa+Ba^\dagger$ and $j^0 = Ca+Da^\dagger$. Using $a|\Omega\rangle = 0$, their equal-time correlators satisfy,
\begin{equation}
    \langle j^0j^0{}^T\rangle  = CD^T = (CB^T)(AB^T)^{-1}(AD^T),
\end{equation}
which gives the relation:
\begin{equation}\label{eq:jtopi_general}
     \langle  j^0_a(\vec{k})j^0_b(\vec{k'})\rangle = \langle  j^0_a(\vec{k})\pi^c(\vec{k'})\rangle \langle \pi(\vec{k}) \pi(\vec{k'})\rangle^{-1}_{cd} \langle \pi^d(\vec{k}) j^0_b(\vec{k'})\rangle
\end{equation}
For type-B NGBs without a quadratic time-derivative term, the Goldstone correlator matrix may be non-invertible. In such cases, one can introduce a regulator $\frac{\chi}{2}\dot\pi^a\dot\pi^a$ and take $\chi\to 0$ after evaluating the inverse relation.

For type-A NGBs, the leading-order relation $j^0_a(\vec{k}) = \Pi_a(\vec{k})$ and the canonical commutation relations further give 
\begin{equation}\label{eq:jtopi}
   F_{ab} =  2\lim_{k\to 0}\langle  j^0_a(\vec{k})j^0_b(-\vec{k})\rangle = \frac{V^2}{2}\lim_{k\to 0}\langle \pi(\vec{k}) \pi(-\vec{k})\rangle^{-1}_{ab} 
\end{equation}
In the IR limit, $\k \to 0$, the left-hand side approaches the charge two-point correlator, which determines the QFM, and the right-hand side involves the inverse Zamolodchikov metric~\cite{Zamolodchikov:1986gt} defined by the NGB correlator. Concretely, if we define the coset construction~\cite{Coleman:1969sm,Callan:1969sn} in terms of 
\beq
U(\pi_a) = \exp(-i \sum_a \pi^a T_a)\ ,
\eeq
then a constant value $\langle \pi_a \rangle = \bar \pi_a = \theta_a$ parametrizes the space of vacua. The information metric, which is proportional to the QFM as a function of $\theta$, is
\beq
g_{ab}(\vec \theta) = {\rm Re} \left[\langle \partial_{\theta_a} \psi |\partial_{\theta_b} \psi \rangle - \langle \partial_{\theta_a} \psi | \psi \rangle \langle \psi | \partial_{\theta_b} \psi \rangle \right] = \frac14 F_{ab}(\vec \theta).
\eeq
For relativistic systems, the metric on the space of vacua (the coset metric) is given by the Zamolodchikov metric via the two-point function 
\beq
\langle \pi^a(\k) \pi^b(-\k)\rangle' = g^{ab}(\pi) /2|\k| \ .
\eeq

where $\langle\ldots\rangle'$ denotes the correlator with delta function omitted.
For non-relativistic theories, there isn't a unique metric. The Lagrangian for Type-A NGBs takes the form
\beq
{\cal L}\supset \frac{1}{2} \bar g_{ab} \dot \pi^a \dot \pi^b - \frac{1}{2}  g_{ab}\partial_i \pi^a \partial^i \pi^b +\ldots \ ,
\eeq
where both $\bar g_{ab}$ and $ g_{ab}$ define ${\cal G}$-invariant metrics~\cite{Watanabe:2012hr,Watanabe:2014fva}. The Zamolodchikov metric and the Fisher metric will be related by
\beq
\langle \pi^a(\k) \pi^b(-\k)\rangle' = \frac{1}{2|\k|} (g\# \bar g)^{-1}\  \ .
\eeq
where $A \# B=A^{1 / 2}\left(A^{-1 / 2} B A^{-1 / 2}\right)^{1 / 2} A^{1 / 2}$ is the matrix geometric mean. The Fisher information metric is then given by $(g\# \bar g)_{ab}$ using \cref{eq:jtopi} at leading order.

The information metric~\cite{Miyaji:2015woj,Stout:2022phm} and coset metric both play key roles in high energy physics. The connection offers the opportunity to apply information-theoretic intuition to effective field theories~\cite{Lashkari:2015hha,Green:2023ids}, or vice versa~\cite{Metlitski:2011pr,Faist:2019ahr}.

\prlsection{Examples} 
As the simplest example, we verify the type-A relation, \cref{eq:jtopi}, in the $O(N)\to O(N-1)$ model,
\begin{equation}
    H = \int d^d x\,\Big[\tfrac{1}{2}\Pi_i\Pi_i+\tfrac{1}{2}\vec \nabla\phi_i \cdot \vec \nabla\phi_i+V(\phi_j)\Big] \ ,
    \label{eqn:ON_model}
\end{equation}
where $\Pi_i$ is the canonical conjugate of $\phi_i$. The vacuum is chosen along the $N$th direction, $\langle \phi_i\rangle = v\delta_{iN}$. The broken charges are therefore
\begin{equation}
    Q_a  =\int d^dx \,[\phi_a(x)\Pi_N(x)-\phi_N(x)\Pi_a(x)]
\end{equation}
Expanding about the symmetry-breaking vacuum to quadratic order and canonically quantizing the fields, we obtain
\begin{equation}
    \begin{aligned}
        F_{ab} = 2\langle \{\Delta Q_a, \Delta Q_b\}\rangle = v^2V  k_{\rm IR} \,\delta_{ab} \ .
    \end{aligned}
\end{equation}
Here we place the system in a periodic box of linear size $L$, with $k_{\rm IR } = \frac{2\pi}{L}\to 0$ in the infinite-volume limit and $V = L^d = \frac{(2\pi)^d}{k_{\rm IR }^d}$.  
Defining the NGB coordinates by  $\phi_a = -v\pi^a+\mathcal{O}(\pi^2)$, the equal-time correlator is  $\langle \pi^a(k)\pi^b(k')\rangle = \frac{(2\pi)^d\delta^{d}(k+k')}{2v^2|k|}\delta^{ab}$, which gives $\frac{V^2}{2}\langle \pi \pi \rangle^{-1}_{ab} = F_{ab}$, verifying~\cref{eq:jtopi}.

Another example is the type-B NGB in a Bose-Einstein condensate with the ferromagnetic ground state. Following~\cite{Watanabe:2012hr,Watanabe:2014fva}, the Lagrangian is given by 
\begin{equation}
    \mathcal{L} = i \psi^\dagger \dot \psi - \frac{1}{2m}\partial_r\psi^\dagger\partial_r\psi+\mu\psi^\dagger\psi -\frac{\lambda}{4}(\psi^\dagger\psi)^2-\frac{\kappa}{4}|\psi^T\psi|^2
    \label{eqn:example_BEC}
\end{equation}
and the ground state is $\psi_{f} = \frac{v_f}{\sqrt{2}}(1,i,0)^T$, where $v_f$ is determined by $\mu = \frac{\lambda v_f^2}{2}$. We can expand the field around the ground state as $ \psi(x) = (v_f + \delta \psi_+(\vec{x})) e_++\delta \psi_-(\vec{x}) e_-+\delta\psi_0(\vec{x}) e_0$, where $e_{\pm}\equiv\frac{1}{\sqrt{2}}(1,\pm i,0)^T$ and $e_0\equiv (0,0,1)^T$. The $\delta\psi_0$ sector decouples at quadratic order and contains two broken charges and the type-B mode. 
\begin{equation}
    Q_1(\vec{k)}  = -\int d^{d}x  \,\frac{v_f\,e^{i\vec{k}\cdot\vec{x}}}{\sqrt{2}}(\delta \psi_0^\dagger+\delta \psi_0) +\mathcal{O}(\delta\psi^2)
\end{equation}
\begin{equation}
    Q_2(\vec{k})  = -i\int d^{d}x  \,\frac{v_f\,e^{i\vec{k}\cdot\vec{x}}}{\sqrt{2}}(\delta \psi_0^\dagger-\delta \psi_0) +\mathcal{O}(\delta\psi^2)
\end{equation}
The Fisher matrix is 
\begin{equation}
    F_{ab}+iU_{ab} = 4\langle \Delta Q_a\Delta Q_b\rangle =     v_f^2V
    \begin{bmatrix}
        1 &i\\
        -i & 1
    \end{bmatrix}
\end{equation}
Defining the corresponding NGB coordinates $\pi^{1,2}$ from $\delta\psi_0$, we find 
\begin{equation}
    \langle \Delta Q_a\Delta Q_b\rangle  = 2\langle \Delta Q_a \pi^c \rangle \langle \pi\pi\rangle^{-1}_{cd}\langle\pi^d \Delta Q_b\rangle
\end{equation}
verifying eq.~\ref{eq:jtopi_general} in the zero-momentum limit. The factor of 2 arises from the Hermitian averaging of $\Delta Q$.  The antisymmetric part reflects the non-commutativity of the two broken charges.

\prlsection{Finite Temperature and General Density Matrices}
For general mixed states, the quantum information geometry is not unique~\cite{Petz:1999xrh}, while for a fixed theory and density matrix, the NGB two-point function determines a definite Zamolodchikov metric. The question is therefore which, if any, mixed-state quantum information metric reproduces its inverse. 
For a type-A NGB in the $O(N)\to O(N-1)$ model at finite temperature, the QFM defined with the SLD is~\cite{Hauke:2015knb} 
\begin{equation}
    F_{ab} = \frac12\text{Tr}(\rho \{L_a,L_b\}) = v^2\omega_{k_{\rm IR}}\tanh(\frac{\beta \omega_{k_{\rm IR}}}{2})\delta_{ab}V
\end{equation}
where $\omega_{k_{\rm IR}} = |k_{\rm IR}|$. Taking $k_{\rm IR}\to 0$, with $V = \frac{(2\pi)^d}{k_{\rm IR}^d}$, the QFM scales as $F_{ab}\sim  k_{\rm IR }^{2-d}\delta_{ab}$, and therefore does not grow with volume when $d\leq2$, consistent with the Mermin-Wagner theorem.
Meanwhile, the NGB correlator is 
\begin{equation}
    \begin{aligned}
        \langle \pi^a(\vec{k}) \pi^b(-\vec{k})\rangle = \frac{\coth(\frac{\beta\omega_k}{2})}{2v^2 \omega_k}\delta^{ab}V \ ,
    \end{aligned}
\end{equation} 
which reproduces the relation $F_{ab} = \frac{V^2}{2} \langle \pi\pi\rangle^{-1}$.

However, this agreement is not generic. The derivation of \cref{eq:jtopi_general} relies on $a|\Omega\rangle = 0$, which does not generally hold for mixed states. To see this, let $\pi = A a+Ba^\dagger$ and an operator $\O = Ca+Da^\dagger$,  and denote $\langle a^2\rangle \equiv m $, $\langle a^\dagger{}^2\rangle \equiv m^*$, and $\langle a^\dagger a\rangle\equiv \bar n$. Their 2-point correlators satisfy  
\begin{equation}
   \langle \O \O\rangle \langle \pi\pi\rangle - \langle \O \pi\rangle \langle \pi \O \rangle  = (AD-BC)^2(|m|^2-\bar n(\bar n +1))
\end{equation}
The inverse relation thus requires either $(|m|^2-\bar n(\bar n +1)) = 0$ or $AD-BC = 0$. Type-A in a thermal state is an example of the latter for $\O=L$. Writing $\pi = Xa+X^*a^\dagger$ and  $Q = i\lambda\Big(Xa-X^*a^\dagger\Big)$, and using
\begin{equation}
    [\rho, a ] = \tanh(\frac{\beta \omega}{2})\{\rho,a\},\; [\rho, a^\dagger ] = -\tanh(\frac{\beta \omega}{2})\{\rho,a^\dagger\},
\end{equation} 
the SLD equation $-i[Q,\rho] = \frac12\{L,\rho\}$ gives 
\begin{equation}
    L = -2\lambda\tanh(\frac{\beta\omega}{2})\Big[Xa+X^*a^\dagger\Big]\propto \pi.
\end{equation} 
On the other hand, for type-B NGBs, the field and the charge may take the form 
\begin{equation}
    \begin{aligned}
        &\pi^1 = X\Big[b_++b_+^\dagger+b_-+b_-^\dagger\Big]\\
        &Q_1 = iY_+\Big[b_+-b_+^\dagger\Big]+iY_-\Big[b_--b_-^\dagger\Big]
    \end{aligned}
\end{equation}
where $X, Y_{\pm}$ are generally distinct coefficients, and $b_{\pm}^\dagger, b_{\pm}$ are the creation and annihilation operators. In this case, $L_1 = -2\Big[\tanh(\frac{\beta\omega^+}{2})Y_+\Big(b_++b_+^\dagger\Big)+\tanh(\frac{\beta\omega^-}{2})Y_-\Big(b_-+b_-^\dagger\Big)\Big]\not\propto\pi^1$. We note, however, that if the gapped $b_+$ mode is integrated out ($\chi=0$), $L_1\propto \pi^1$ , and the inverse relation is recovered in the resulting thermal type-B theory.

In general, open systems have both strong and weak forms of spontaneous symmetry breaking~\cite{Albert:2014yej,Hongo:2019qhi,Lessa:2024gcw,Hauser:2026sgr}, and it will be interesting to see if the difference can be characterized via Fisher information following our approach.

\prlsection{Conclusions}\label{sec:conclusions} Understanding the unique role of quantum mechanics in the Universe is one of the central organizing questions of modern physics~\cite{Faulkner:2022mlp}. Whether we are trying to understand the origin of quantum advantages~\cite{Harrow:2017jkm} or origin of structure of the Universe~\cite{Maldacena:2015bha,Green:2020whw}, we have yet to fully characterize when and how quantum evolution leads to measurably distinct outcomes from their classical counterparts. Naturally, information-theoretic quantities, like the quantum Fisher information, characterize the measurable content of these states and can expose the underlying quantum nature of these systems. The challenge is that these quantities are static, and so it takes work to determine how they are influenced by the dynamics, or more precisely, the Hamiltonian~\cite{Wan:2021enx}. 

In this paper, we offered a new connection between the Fisher information of the ground state of a quantum system and the dynamics encoded in NGBs. The properties of NGBs have been well studied, particularly using effective field theory, and provide new intuition for the structure of quantum information. The connection between the quantum aspects of the Fisher information associated with the Uhlmann curvature and the charge density matrix characterizing Type B Goldstones gives us a new language and physical intuition for when quantum effects are essential in measurement of complex systems. We hope this will offer new insights into the quantum nature of quantum information, and how to probe quantum dynamics observationally~\cite{Watts:2021thg,Green:2022fwg,Green:2026edk}.

\begin{acknowledgments}
We are grateful to JJ Carrasco, Peter Graham, Kshitij Gupta, Austin Joyce, Aneesh Manohar, John McGreevy, Akhil Premkumar, Nicholas Rodd, and Eva Silverstein.
We are supported by the US~Department of Energy under grant~\mbox{DE-SC0009919}. 
\end{acknowledgments}

\phantomsection
\addcontentsline{toc}{section}{References}
\small
\bibliographystyle{utphys}
\bibliography{Refs}

\clearpage
\onecolumngrid
\begin{center}
{\large \bf Supplemental Material}
\end{center}

\setcounter{equation}{0}
\setcounter{section}{0}
\setcounter{figure}{0}
\setcounter{table}{0}
\renewcommand{\theequation}{S\arabic{equation}}
\setcounter{secnumdepth}{3}
\renewcommand{\thesection}{S\Roman{section}}
\renewcommand{\thefigure}{S\arabic{figure}}
\renewcommand{\thetable}{S\arabic{table}}

\section{Nambu-Goldstone Boson Conventions}\label{app:goldstone}

The broken symmetries can be characterized by the expectation value of operators charged under the symmetry. Suppose that we have some scalar operator $\phi_\ell$, where $\ell$ labels some $M$-dimensional representation of ${\cal G}$, that transforms linearly under the symmetry and acquires a vacuum expectation value
\beq
\langle \phi_\ell \rangle = v_\ell \to \langle [Q_a, \phi_\ell ] \rangle = -i T^a_{\ell\ell'} v^{\ell'}
\eeq
To make the counting of broken generators straightforward, we define an operator 
\beq
X^a =\phi^\ell T^a_{\ell \ell'}v^{\ell'} 
\eeq
In this basis, the commutator with the charge is
\beq
\langle [Q_a , X_b ]\rangle = -i v^\ell T^a_{ m \ell}T^b_{m\ell'} v^{\ell'} =- i \kappa_{ab}   \ .
\eeq
Notice that $\kappa_{ab}$ is a positive symmetric matrix. As a consequence, we can always locally choose a basis where it is diagonal, and can further rescale $X_a$ to make $\kappa_{ab} =\delta_{ab}$ (this cannot be done globally as a function of $v^\ell$).

Now let's move on to the currents and the implications of the states of the theory. The charges defined above are integrals of the currents so that $Q_a = \int d^d x j^0_a(\x,t)$ and therefore, using causality, we have
\beq
\langle [j^0_a(\x,t=0), X_b(0) ] \rangle = -i \kappa_{ab} \delta(\x) \ .
\eeq
Inserting a complete set of states shows the existence of the Nambu-Goldstone boson,
\beq
\int d^d x\langle j^0_a(\x,t) X_b(0) \rangle =\int d^d x \sum_{n,\p} \langle j^0_a(\x=0,t=0) e^{-i E_n t} e^{i \p \cdot \x} |n, \p\rangle \langle n, \p| X_b(0) \rangle \ .
\eeq
Since $\int d^d x j^0_a(\x,t) = Q_a$ is conserved, this correlation must be time-independent. Integrating over $\x$ gives $\p =0$ and the time-independence gives us $E_n \to 0$. For relativistic theories, one can go further to show this contains a single-particle state $|\pi, \p\rangle$ so that at equal time $t=0$, we define
\beq
\langle j_a^0 (\x)|\pi_c,\p\rangle =  e^{i\p\cdot\x}{\cal J}_{ac} (\p) \qquad \langle j_a^i (\x)|\pi_c,\p\rangle = e^{i\p\cdot\x}{\cal J}^i_{ac}  (\p) \qquad \langle X_b(0) |\pi_c,\p \rangle  = \Phi_{bc} (\p)
\eeq
so that 
\beq
\kappa_{ab} = -2{\rm Im}  \sum_c {\cal J}_{ac} \Phi^{\dagger}_{cb} \ .
\eeq
From here, we can also define the current-current two-point statistics
\beq
\langle j_a^0(\x) j_b^0(0)\rangle = \int \frac{d^d p}{(2\pi)^d} e^{i\p\cdot x} \sum_c {\cal J}_{ac} {\cal J}^\dagger_{c b} 
\eeq
Moreover, since $\langle j^0_a (\x)\rangle= 0$ in a Lorentz invariance vacuum, $\langle [j^0_a,j^0_b ] \rangle = 0$ and the RHS is both real and positive semidefinite.

We are interested in understanding non-relativistic theories as well and therefore we need to include a generalization of the above to vacuum states of non-relativistic QFTs. In order to simplify our discussion, we will define generalization of the charges with momenta, as follows
\beq
Q_a(\k) = \int_V d^dx e^{i\k\cdot \x} j^0_a(\x) \ ,
\eeq
where $V$ is the volume over which the integral is performed. We will anticipate taking $k\to 0$ and $V\to \infty$ so that $Q_a(\k)\to Q_a$ but it will be technically and conceptually useful to maintain both at finite values for most calculations. We then define matrix elements as
\bea
{\cal J}_{ab}(\k) &=& \langle j_a(0) |\pi_b, \k\rangle \to \langle Q_a(\k') |\pi_b, \k\rangle = (2\pi)^d\delta^{(d)}(\k'-\k) {\cal J}_{ab}(\k)  \\
 {\cal J}^i_{ab}(\k) &=& \langle j_a^i(0) |\pi_b, \k\rangle  \qquad  \Phi_{ab}(\k) = \langle X_a  |\pi_b, \k\rangle.
\eea
Current conservation then relates $E(\k){\cal J}_{ab}(\k) = k_i {\cal J}^i_{ab}(\k)$. Now we can define the finite-momentum versions of our charge correlators

\begin{equation}
\begin{aligned}
    & \langle [Q_a(\k), X_b(0) ] \rangle = - i \kappa_{ab} = 2i{\rm Im} {\cal J}_{ac} (\k)\Phi^\dagger_{cb}(\k) \\\
    &\langle Q_a(\k) Q_b(-\k) \rangle = V\,\text{Re}( {\cal J}_{ac} {\cal J}^\dagger_b{}^c) +\frac{ i V}{2} \varrho_{ab} \ ,
\end{aligned}
\end{equation}
when $\langle Q_a\rangle = 0$, the second equation can be written as 
\begin{equation}
    \langle Q_a(\k) Q_b(-\k) \rangle = \frac12 F_{ab}+\frac{ i V}{2} \varrho_{ab} 
\end{equation}
where we are using the replacement $(2\pi)^d \delta^{(d)}(0) \to V$ to account for the finite volume.

\subsection{Nambu-Goldstone Boson and the Noether Current from the EFT}
Assuming spatial rotational symmetry, the general form of NGB's EFT up to the second order in derivatives is 
\begin{equation}
    \mathcal{L}_{\rm eff} = c_a(\pi)\dot\pi^a+\frac12 \bar{g}_{ab}(\pi)\dot \pi^a \dot \pi^b - \frac12 g_{ab} (\pi) \vec{\nabla}\pi^a\cdot \vec{\nabla}\pi^b
\end{equation}
The definition of the conjugate momentum of $\pi^a $ is 
\begin{equation}
    \Pi_a(x) = \frac{\partial S_{\rm eff}}{\partial \dot\pi^a(x)}
\end{equation}
where the functional derivative $\frac{\partial f(y)}{\partial f(x)} \equiv \delta^{(d+1)}(x-y)$.
The Noether current 
\begin{equation}
    j^\mu_a(x) = \frac{\partial S_{\rm eff}}{\partial(\partial_\mu\pi^b(x))}\xi^b_a(\pi) - K^{\mu}_a(x)
     \label{eqn:noether_current}
\end{equation}
has to match the Noether current in the UV theory. Here $K^\mu_a$ is defined through $\delta_\epsilon \mathcal{L}_{\rm eff} = \partial_\mu (\epsilon^a K^\mu_a)$ , so the action is still invariant under $\pi^a \to \pi^a+\epsilon^b\xi_b^a(\pi)$, where $\xi^a_b(\pi)\equiv \delta_b^a+\mathcal{O}(\pi)$ is the Killing vector. Matching up the current in the UV theory and the IR theory uniquely defines the Goldstone fields $\pi^a$.
If we expand the Lagrangian around the vacuum, with $G_{ab}\equiv \bar{g}_{ab}(0)$, $\rho_{ab}=-\rho_{ba}$,
\begin{equation}
    \mathcal{L}  = \frac12 G_{ab}\dot\pi^a\dot\pi^b -\frac12 \varrho_{ab}\pi^a\dot\pi^b  - \frac12 g_{ab}\partial_i\pi^a\partial_i\pi^b+\mathcal{O}(\pi^3 )
\end{equation}
since $\delta_{\epsilon}\mathcal{L} = -\epsilon^a\frac12\rho_{ab}\dot \pi^b$, $K^0_a = -\frac12 \rho_{ab}\pi^b$. And 
\begin{equation}
    \begin{aligned}
        j^0_a = \Pi_a +\frac12\varrho_{ab}\pi^b +\mathcal{O}(\pi^2)
    \end{aligned}
\end{equation}
In the IR limit, it is the charge
\begin{equation}
    Q_a \equiv \int d^{d}x j^0_a(x) = \lim_{k_{\rm IR}\to 0}j^0_a(k_{\rm IR}) \ .
\end{equation}
The nuance in the theory of NGBs has precisely to do with this limit. However, it is important that we are assuming $k_{\rm IR}$ is much smaller than the other scales in the problem.

\section{Quantum Fisher Matrix Conventions}\label{app:fisher}

\subsection{Generalities}

We will follow the definition of the quantum Fisher matrix in terms of the symmetric logarithmic derivative (SLD) of the density matrix, $L_i$, associated with a parameter $\theta_i$ via
\beq
\frac{\partial}{\partial \theta_i} \rho =\frac{1}{2} \{ L_i ,\rho \}  
\eeq
Here $L_i$ is a Hermitian operator and therefore cannot be simply expressed in terms of $\partial_\theta$ alone, but instead depends on the density matrix.
Note that for a classical density matrix, 
\beq
\rho = \delta_{ab} P_a(\vec \theta) \to L_i = \delta_{ab} \frac{\partial}{\partial \theta_i} \log P_a(\vec \theta) \ .
\eeq
This will provide some useful intuition when comparing the classical and quantum versions for the Cram\'er-Rao bound.

Given the SLD, we define the quantum Fisher matrix as 
\beq
F_{ab} = \frac{1}{2} {\rm Tr} \rho \{ L_a ,L_b\}
\eeq
The utility of this construction can be seen through its similarity to the classical Fisher information and the Cram\'er-Rao bound. Suppose we define a quantum estimator for $\theta_i$, $\hat \theta_i$, and then require that it is unbiased so that 
\beq
{\rm Tr} \rho \hat \theta_i = \theta_i \quad \to \quad \partial_{\theta_j}{\rm Tr} \rho  \hat \theta_i ={\rm Tr}( \partial_{\theta_j} \rho) (\hat \theta_i - \theta_i) = \delta_{ij} \ ,
\eeq
where in the last step we used ${\rm Tr}\partial_{\theta_i} \rho = 0$ to subtract the mean. Now we notice that we can write this expression in terms of the SLD,
\beq
\frac{1}{2} {\rm Tr} \{L_j ,\rho\} (\hat \theta_i - \theta_i)  =\frac{1}{2}  {\rm Tr}\rho \{L_j ,  (\hat \theta_i - \theta_i) \} = \delta_{ij} \ .
\eeq
From here it is easy to see the origin of the quantum Cram\'er-Rao bound. We notice that we can define a positive symmetric matrix
\beq
\Delta_{ij} = \frac{1}{4} {\rm Tr}\rho \{ L_i -  t(\hat \theta_i - \theta_i), L_j - t(\hat \theta_j - \theta_j) \}   \geq 0 \ .
\eeq 
where $t$ is an arbitrary real number.
Minimizing with respect to $t$ implies the bound
\beq
F_{ii} {\rm Tr} \rho (\hat \theta_i - \theta_i)^2 \geq \left(\frac{1}{2} {\rm Tr}\rho \{L_i,  (\hat \theta_i - \theta_i) \} \right)^2 = 1 \ .
\eeq
Without even worrying about quantum mechanics, one can recognize that the Cram\'er-Rao bound should involve the matrix inverse. However, we notice that if 
\beq
\Sigma_{ij} = \frac{1}{2} {\rm Tr} \rho \{ (\hat \theta_i - \theta_i), (\hat \theta_j - \theta_j) \}  \quad \to \quad G=\left(\begin{array}{ll}
\Sigma & I \\
I & F
\end{array}\right) \geq 0 \ ,
\eeq
where positivity of $G$ follows from the positivity of $\Delta_{ij}$. Since $G$ is positive, we can change basis and it would still be positive and therefore
\beq
S = \left(\begin{array}{cc}
I & 0 \\
-F^{-1} & I
\end{array}\right) \to S^{T} G  S = \left(\begin{array}{ll}
\Sigma - F^{-1} & 0\\
0 & F
\end{array}\right) \geq 0 \quad  \to \quad \Sigma \geq F^{-1} \ .
\eeq
This is the quantum Cram\'er-Rao bound.

One unexpected feature of the QCR bound is that it doesn't seem to involve any non-zero commutators. This is a signal that the QCR bound is not necessarily optimal, particularly for variables that don't commute. Specifically, if we define the Uhlmann curvature
\beq\label{eq:Us}
U_{ab} = \frac{1}{2i} {\rm Tr} \rho[L_a, L_b] \ ,
\eeq
then the Cram\'er-Rao bound can only be saturated when $U_{ab} = 0$. In practice, the most general bound we can saturate is the Holevo-Cram\'er-Rao~\cite{Holevo:2011} bound that can be defined by minimizing over all unbiased estimators. Specifically, given any unbiased estimators $\hat X_i =\hat \theta_i - \theta_i$ satisfying, 
\beq
{\rm Tr}\partial_{\theta_i} \rho \hat X_j = \delta_{ij} \ ,
\eeq
we can calculate the matrix
\beq
Z_{ij} = {\rm Tr} \rho \hat X_i \hat X_j  \ .
\eeq 
The Holevo-Cram\'er-Rao bound then states that the minimum variance measurement that is achievable for a positive operator valued measure is 
\beq
{\rm Tr} W \Sigma \geq {\rm min}_{X} \left[ {\rm Tr} W {\rm Re} Z +|| W^{1/2} {\rm Im} Z W^{1/2} || \right] \ ,
\eeq
where the minimization is over all unbiased estimators $\hat X$ and $||..||$ is the trace norm. We can often take the optimal estimator to be
\beq\label{eq:estimatorX}
\hat X = F^{-1}  L - \langle \hat X\rangle
\eeq
such that 
\beq
{\rm Tr} W \Sigma \geq  \left[ {\rm Tr} W F^{-1} +|| W^{1/2} F^{-1} U F^{-1} W^{1/2} || \right] \ ,
\eeq
Let's see why this is often the correct choice. Suppose we are wrong and another estimator, 
\beq
 X' = F^{-1}  L +  K - \langle \hat X\rangle  \ ,
\eeq 
gives a smaller value. Since $\hat X'$ must still be an unbiased estimator, we  have 
\beq
\frac12{\rm Tr} \rho \{L_a, X_b\} = \delta_{ab} \to {\rm Tr}\rho \{ L_a, K_b \} = 0 \ .
\eeq
This implies that the change to the real part of $Z$ is positive,
\beq
{\rm Re Z} = F^{-1} + F^{-1} \frac{1}{2} {\rm Tr} \rho \{L_a,K_b \} + \frac{1}{2} {\rm Tr} \rho \{K_a,L_b \} F^{-1}+\frac{1}{2} {\rm Tr}\rho\{K_a,K_b\} = F^{-1}+\frac{1}{2} {\rm Tr}\rho\{K_a,K_b\}
\eeq
The real part of this expression is therefore minimized by the estimator in \cref{eq:estimatorX}.

The imaginary part is less straightforward to understand. Since the norm is manifestly positive, when $U_{ab}= 0$ we cannot lower the bound further and this is the optimal choice and reproduces the Cram\'er-Rao bound. When $U_{ab} \neq 0$, it is possible to find operators $K_a$ that lower the bound. Defining $N = W^{1/2} F^{-1} U F^{-1} W^{1/2} \neq 0$, we need ${\rm Tr}\sqrt{ N^\dagger N}\neq 0$ in order to have a linear term in $K_a$ that can take either sign. Defining
\beq
A_{ab} =\frac{1}{2i} \left({\rm Tr}\rho [X_a, K_b] + {\rm Tr}\rho [K_a, X_b]\right) \ ,
\eeq
it is possible to find a lower variance estimator if there is an operator $K_a$ that gives rise to $A_{ab} \neq 0$. We will see when this is possible for the case of pure states.


\subsection{Pure States and Symmetry Breaking}

Now suppose that we want to measure some set of parameters $\theta_a$ defined by the state
\beq
|\Psi \rangle = e^{ -i \sum_a\theta^a Q_a } |\Omega \rangle  \ ,
\eeq
where $Q_a$ is a zero-momentum charge. The symmetric  logarithmic derivative is defined by
\beq
L_a  = -2i [Q_a,\rho] \to F_{ab} = \frac{1}{2} {\rm Tr} \rho \{L_a, L_b \} = 2\langle \{ \Delta Q_a , \Delta Q_b \} \rangle
\eeq
where $\Delta Q_a = Q _a -\langle Q_a \rangle$. In practice, we will use the finite momentum ``charges" so that $Q_a \to \frac{1}{2} (Q_a(\k)+Q_a(-\k))$ so the operator is Hermitian. This will leave the form of the QFM the same if we define
\beq
\Delta Q_a \to \frac{1}{2} (Q_a(\k)+Q_a(-\k)) -\ \frac{1}{2} (\langle Q_a(\k)\rangle+\langle Q_a(-\k)\rangle) \ .
\eeq
Technically speaking, at finite momentum $\langle Q_a(\k)\rangle$ should vanish, by translation invariance either at $V \to \infty$ or when placed in a periodic box. However, general boundary conditions for finite volume will break translation invariance and so we will keep these terms for completeness. 

Using the definition of the Uhlmann curvature in \cref{eq:Us}, we have
\beq
U_{ab} =\frac{1}{2i} {\rm Tr} \rho [L_a,L_b] = \frac{2}{i} \langle [ \Delta Q_a, \Delta Q_b]  \rangle \ .
\eeq
The non-trivial question is what is the Holevo-Cram\'er-Rao bound when $U_{ab} \neq 0$, as it is possible to find estimators that give lower variance than \cref{eq:estimatorX}. For a pure state, we can see that for $A_{ab}$ to be non-zero, it requires $K_b$ to be related to $i L_a$ without being in the span of $L_a$~\cite{Matsumoto:1997hd}. Specifically, if we define $K_a |\psi\rangle  = |k_a\rangle$ and $X_b |\psi \rangle = |x_b\rangle$ then
\beq
{\rm Tr} \rho\{K_a, \hat X_b\} = 2{\rm Re}\langle k_a| x_b \rangle = 0 \qquad  {\rm Tr} \rho [K_a, \hat X_b] = 2i {\rm Im}\langle k_a| x_b \rangle \neq 0 \ .
\eeq
In other words, we need $K_a$ to produce a state with overlap with $\hat X$.

A concrete realization of this possibility is $K_a = \Delta Q_b$. Since $L_b = -2i [\Delta Q_b, \rho]$, these generate related states but we have 
\beq
{\rm Tr} \rho\{K_a, L_b\} = - 2i \langle [ \Delta Q_b ,\Delta Q_b ] \rangle  = 0 \qquad {\rm Tr} \rho [K_a, L_b] =-4i \langle  \Delta Q_b ^2 \rangle  \neq 0 \ .
\eeq
Of course, this can only lower the bound if the charge $\Delta Q_b$ is type A (the state would be in the span of $L_a$ is it has a type B partner). At the same time, we need $\hat X_a$ to be associated with type B so that it has a non-zero $U_{ab}$. Both conditions must be present to lower the bound.

\section{Matching}\label{app:matching}
\subsection{Inverse relation between charge and Goldstone Boson correlators}
Intuitively, the symmetry of a state implies indistinguishability under the corresponding symmetry transformation, whereas quantum information measures distinguishability. This motivates the use of quantum information to characterize symmetry breaking~\cite{Zanardi:2007mqd,Venuti:2007izd,Gu:2008dyh,Kolodrubetz:2013mla,Ares:2022koq,Yamashika:2025ghp,Yamaguchi:2026mqp,simeth2026}. One goal of this paper is to establish the operational relation between symmetry and information.

When we express the Noether currents and the NGB fields in terms of the creation and annihilation operators
\begin{equation}
    \begin{aligned}
        \pi^a(\vec{k}) = A^a_b(\vec{k}) a_{k}^b+ B^a_b(\vec{k}) (a_{k}^\dagger)^b\\
        (j^{0})_a(\vec{k}) = C_{ab}(\vec{k}) a_{k}^b+ D_{ab}(\vec{k}) (a_{k}^\dagger)^b
    \end{aligned}
\end{equation}
The equal-time two point correlators are thus
\begin{equation}\label{eqn:pure_state_correlators}
    \begin{aligned}
       & \langle \pi^a(\vec{k}) \pi^b(\vec{k'})\rangle = A^a_c (\vec{k})B^{cb}(\vec{k'})\;(2\pi)^d\delta^{(d)}(\vec{k}+\vec{k'})\\
        &\langle \pi^a(\vec{k}) j^0_b(\vec{k'})\rangle = A^a_c (\vec{k})D^{c}_b(\vec{k'})\;(2\pi)^d\delta^{(d)}(\vec{k}+\vec{k'})\\
        &\langle  j^0_a(\vec{k})\pi^b(\vec{k'})\rangle = C_{ac} (\vec{k})B^{cb}(\vec{k'})\;(2\pi)^d\delta^{(d)}(\vec{k}+\vec{k'})\\
        &\langle j^0_a(\vec{k})j^0_b(\vec{k'})\rangle = C_{ac} (\vec{k})D^{c}_b(\vec{k'})\;(2\pi)^d\delta^{(d)}(\vec{k}+\vec{k'})
    \end{aligned}
\end{equation}
Thus a simple relation follows:
\begin{equation}
    \langle  j^0_a(\vec{k})j^0_b(-\vec{k})\rangle = \langle  j^0_a(\vec{k})\pi^c(-\vec{k})\rangle \langle \pi(\vec{k}) \pi(-\vec{k})\rangle^{-1}_{cd} \langle \pi^d(\vec{k}) j^0_b(-\vec{k})\rangle
\end{equation}
For type-A NGB, this relation can be further simplified, since the zeroth component of the current is just the canonical conjugate of the NGB to the leading order $j^0_a = \Pi_a$. With the commutation relation of canonical quantization $[\pi^a(\vec{k}), \Pi_b(\vec{k'})] = i\delta^a_b (2\pi)^d\delta^{(d)}(\vec{k}+\vec{k'})$ and the fact that $\langle\{\Pi_a,\pi^b\}\rangle = 0$ with a quadratic Goldstone EFT, one can further simplify the relation with $\langle  j^0_a(\vec{k})\pi^b(-\vec{k})\rangle  = \frac{-i}{2}\delta^a_b (2\pi)^d\delta^{(d)}(0)$
\begin{equation}
    \langle  j^0_a(\vec{k})j^0_b(-\vec{k})\rangle = \frac{V^2}{4}\langle \pi(\vec{k}) \pi(-\vec{k})\rangle^{-1}_{ab} 
    \label{eqn:jj-pipi_relation}
\end{equation}
where $V$ is the spatial volume of the periodic box, with $(2\pi)^d\delta^{(d)}(0) = V$. 

\subsection{Type-A Example}
Consider the $O(N)\to O(N-1)$ model

\begin{equation}
    L = \int d^dx \Big[\frac12 \partial_\mu\phi_i(x)\partial^\mu\phi_i(x) -V(\phi_i)\Big]
    \label{eqn:ON_model_Lagrangian}
\end{equation}
\begin{equation}
    H = \int d^d x\,\Big[\frac{1}{2}\Pi_i(x)\Pi_i(x)+\frac{1}{2}\nabla\phi_i(x)\nabla\phi_i(x)+V(\phi_j)\Big]
\end{equation}
Choosing the vacuum direction to be $N$, such that $\langle \phi_i \rangle = v\delta_{iN} $,
\begin{equation}
    Q_a (\vec{k}) =\int d^dx\, e^{i\vec{k}\cdot\vec{x}}\,[\phi_a(x)\Pi_N(x)-\phi_N(x)\Pi_a(x)]\approx -v\int d^dx \,e^{i\vec{k}\cdot\vec{x}}\,\Pi_a(x)
\end{equation}
For type-A NGBs, using the standard canonical quantization of the quadratic Goldstone Hamiltonian, each momentum mode is a harmonic oscillator,
\begin{equation}
    \Pi_a(k) = i\sqrt{\frac{\omega_k}{2}}[a^a_{k}e^{-i\omega_k t}-(a^\dagger_{k})^ae^{i\omega_k t}]
\end{equation}
Taking the expectation value in the oscillator ground state gives
\begin{equation}
    \begin{aligned}
        &\langle \Pi_a(k)   \,\Pi_b(k')\rangle 
        =(2\pi)^d\delta^{(d)}(k+k')\frac{\omega_k}{2}\delta_{ab}
    \end{aligned}
\end{equation}
where $\omega_k = |k|$. And $[a^a_k,(a^\dagger_{k'})^b] = \delta^{ab (2\pi)^d }\delta^{(d)}(\vec{k}+\vec{k'})$
Therefore, the QFM is 
\begin{equation}
    \begin{aligned}
        F_{ab} = 2\langle \{\Delta Q_a, \Delta Q_b\}\rangle = v^2V  k_{\rm IR} \,\delta_{ab}
    \end{aligned}
\end{equation}
where $V = \Big[\frac{2\pi}{k_{\rm IR}}\Big]^d$ is the volume of the system with periodic boundary conditions, and $k_{\rm IR} \to 0$ is taken to recover the total charge.
To compare it with the 2-point statistics of the NGBs, we first need to write down its EFT, with $\phi_a = -v\pi^a+\mathcal{O}(\pi^2)$

\begin{equation}
    L_{\rm EFT} = \int d^dx \Big[\frac{v^2}{2}\delta_{ab}\partial_\mu\pi^a\partial^\mu\pi^b \Big] +\mathcal{O}(\pi^3)
\end{equation}
\begin{equation}
    H_{\rm EFT} = \int d^dx \Big[\frac{1}{2v^2}\Pi^{(\pi)}_a(x)\Pi^{(\pi)}_a(x)+\frac{v^2}{2}\nabla\pi^a(x)\nabla\pi^a(x)\Big] +\mathcal{O}(\pi^3)
\end{equation}
The leading term is the familiar simple harmonic oscillator. 
\begin{equation}
    \pi^a(k) = \frac{1}{v\sqrt{2\omega_k}}\Big[a^a_{k}e^{-i\omega_k t}+(a^\dagger_{k})^ae^{i\omega_k t}\Big]
\end{equation}
and the conjugate momentum to $\pi^a$ is
\begin{equation}
    \Pi^{(\pi)}_{a}(k) \equiv j^0_a(k) =  -iv\sqrt{\frac{\omega_k}{2}}[a^a_{k}e^{-i\omega_k t}-(a^\dagger_{k})^ae^{i\omega_k t}]
\end{equation}
They give the 2-point correlator in the following expression
\begin{equation}
    \langle \pi^a(k)\pi^b(k')\rangle = \frac{(2\pi)^d\delta^{d}(k+k')}{2v^2|k|}\delta^{ab}
    \label{eqn:ON_pipi}
\end{equation}
Now we can verify the relation in eq.~\ref{eqn:jj-pipi_relation} 
\begin{equation}
    \begin{aligned}
      & F_{ab} =  2\langle \{\Delta Q_a, \Delta Q_b\}\rangle  =  \langle \Big(j^0_a(\vec{k}_{\rm IR})+h.c.\Big)\Big( j^0_b(-\vec{k}_{\rm IR})+h.c.\Big)\rangle \\
       &= \frac{V^2}{2}\langle \pi(\vec{k}_{\rm IR}) \pi(-\vec{k}_{\rm IR})\rangle^{-1}_{ab} =  Vv^2k_{\rm IR}\delta_{ab}
    \end{aligned}
\end{equation}
This echoes Coleman's theorem: Since $\langle \pi^a(\vec{x})\pi^b(\vec{0})\rangle'= \int \frac{d^dk}{(2\pi)^d}e^{i\vec{k}\cdot \vec{x}}\frac{\delta^{ab}}{2v^2|k|}$ has an IR divergence for spatial dimensions $d\leq 1$, the symmetry breaking is unstable. Here, from the QFM $F_{ab} = \lim_{k_{\rm IR} \to 0} Vv^2k_{\rm IR}\delta_{ab}$, we also see that the information does not grow with system size when the spatial dimension $d\leq 1$. In this sense, the absence of symmetry breaking is reflected in the failure of the QFI growing with system size.

\subsection{Type-B Example}
For type-B NGBs, let us consider a spinor Bose-Einstein condensate with the ferromagnetic ground state as in ~\cite{Watanabe:2012hr}.
\begin{equation}
    \mathcal{L} = i \psi^\dagger \dot \psi - \frac{1}{2m}\partial_r\psi^\dagger\partial_r\psi+\mu\psi^\dagger\psi -\frac{\lambda}{4}(\psi^\dagger\psi)^2-\frac{\kappa}{4}|\psi^T\psi|^2
    \label{eqn:Lag_BEC}
\end{equation}
and the ground state is $\psi_{f} = \frac{v_f}{\sqrt{2}}(1,i,0)^T$where $v_f$ is determined by $\mu = \frac{\lambda v_f^2}{2}$.

This original system has $U(1)\times SO(3)$ symmetry. The corresponding charges are:
\begin{equation}
    Q_{0}  = \int d^{d}x \psi^\dagger\psi
\end{equation}
\begin{equation}
    Q_{i}  = -i\epsilon_{ijk}\int d^{d}x \psi_j^\dagger\psi_k, \, i\in \{1,2,3\}
\end{equation}

 Now we parametrize the full field $\psi$ as 
\begin{equation}
    \psi(x) = (v_f + \delta \psi_+(\vec{x})) e_++\delta \psi_-(\vec{x}) e_-+\delta\psi_0(\vec{x}) e_0
\end{equation}
where $\frac{1}{\sqrt{2}}(1,i,0)^T\equiv e_+$, $\frac{1}{\sqrt{2}}(1,-i,0)^T\equiv e_-$. and $(0,0,1)^T\equiv e_0$.
Then the broken charges can be simplified as follows
\begin{equation}
    \begin{aligned}
        &\Delta Q_0  = \int d^{d}x  \,v_f(\delta \psi_+^\dagger+\delta \psi_+) +\mathcal{O}(\delta\psi^2)\\
   & \Delta Q_1  = -\int d^{d}x  \,\frac{v_f}{\sqrt{2}}(\delta \psi_0^\dagger+\delta \psi_0) +\mathcal{O}(\delta\psi^2)\\
    &\Delta Q_2  = -i\int d^{d}x  \,\frac{v_f}{\sqrt{2}}(\delta \psi_0^\dagger-\delta \psi_0) +\mathcal{O}(\delta\psi^2)
    \label{eqn:type_B_charges}
    \end{aligned}
\end{equation}
And the Hamiltonian can be rewritten as 

\begin{equation}
    H = \int d^{d}x \mathcal{H}_++\mathcal{H}_-+\mathcal{H}_0 +\mathcal{O}(\delta \psi^3)
\end{equation}
with 
\begin{equation}
    \mathcal{H}_+ = \frac{1}{2m}|\nabla\delta\psi_+|^2+\frac{\lambda v_f^2}{4}(\delta\psi_+^\dagger+\delta\psi_+)^2+\text{const.}
\end{equation}
\begin{equation}
    \mathcal{H}_- = \frac{1}{2m}|\nabla\delta\psi_-|^2+\kappa v_f^2|\delta\psi_-|^2
\end{equation}
\begin{equation}
    \mathcal{H}_0 = \frac{1}{2m}|\nabla\delta\psi_0|^2
\end{equation}
so that the three modes $\delta \psi_\pm,\delta \psi_0$ are decoupled. And  the conjugate momentum of $\delta \psi_\alpha$ is its Hermitian conjugate $\Pi_\alpha = i\delta \psi^\dagger_\alpha$. So the commutation relation between $\delta \psi_\alpha $ and $\delta \psi^\dagger_\alpha$ is fixed 
\begin{equation}
    [\delta \psi_{\alpha}(\vec{x},t),\delta \psi_\beta^\dagger(\vec{y},t)] = \delta_{\alpha \beta} \delta^{(d)}(\vec{x}-\vec{y})
\end{equation}
where $\alpha, \beta \in\{ +,-,0\}$.
Next, we need to use the Heisenberg equation to find the creation and annihilation operators.
For the $\delta\psi_0$ mode,
\begin{equation}
    i\dot{\delta\psi_0}(\vec{k},t) = [\delta \psi_0(\vec{k},t),H_0] = \frac{k^2}{2m}\delta \psi_0(\vec{k},t)
\end{equation}
\begin{equation}
    i\dot{\delta\psi_0^\dagger}(\vec{k},t) = [\delta \psi_0^\dagger(\vec{k},t),H_0] = -\frac{k^2}{2m}\delta \psi_0^\dagger(\vec{k},t)
\end{equation}
so $\delta \psi_0(\vec{k},t)  = e^{-i\omega^0_k t}a_k$ and $\delta \psi_0^\dagger(\vec{k},t)  = e^{i\omega^0_k t}a_{k}^\dagger$ with $a_k^\dagger, a_k$ the creation and annihilation operators $[a_k,a_{k'}^\dagger] = (2\pi)^{d}\delta^{(d)}(\vec{k}+\vec{k'})$ and $\omega^0_k\equiv \frac{ k^2}{2m}$.
And for the $\delta \psi_+$ mode, since the Heisenberg equations of $\delta \psi_+$ and $\delta \psi_+^\dagger$ are mixed,
\begin{equation}
    i\dot{\delta\psi_+}(\vec{k},t) = [\delta \psi_+(\vec{k},t),H_+] = \Big[\frac{k^2}{2m}+\mu \Big]\delta \psi_+(\vec{k},t)+\mu\delta \psi_+^\dagger(\vec{k},t)
\end{equation}
\begin{equation}
    i\dot{\delta\psi_+^\dagger}(\vec{k},t) = [\delta \psi_+^\dagger(\vec{k},t),H_+] = -\Big[\frac{k^2}{2m}+\mu \Big]\delta \psi_+^\dagger(\vec{k},t)-\mu\delta \psi_+(\vec{k},t)
\end{equation}
the solution is more complicated.

\begin{equation}
    \delta\psi_+(\vec{k},t) = e^{-i\bar\omega_k t}\cosh\theta_B \,b_k+e^{i\bar\omega_k t}\sinh\theta_B \,b_{k}^\dagger
\end{equation}
\begin{equation}
    \delta\psi_+^\dagger(\vec{k},t) = e^{-i\bar\omega_k t}\sinh\theta_{B} \,b_k+e^{i\bar\omega_k t}\cosh\theta_B \,b_{k}^\dagger
\end{equation}
where $   \theta_B  = -\frac12 \tanh^{-1}\Big(\frac{\mu}{\frac{k^2}{2m}+\mu }\Big)$, and $\bar\omega_k = \sqrt{\frac{k^2}{2m}(\frac{k^2}{2m}+2\mu)}$.
Now,
\begin{equation}
    \begin{aligned}
        & \langle \Delta Q_1 \Delta Q_1\rangle = \frac{v_f^2}{8} \lim_{k\to 0} \langle \Big(\delta \psi_0( \vec{k}) +\delta \psi_0^\dagger( \vec{k}) +h.c.\Big)\Big(\delta\psi_0( -\vec{k})+\delta\psi_0^\dagger( -\vec{k})+h.c.\Big)\rangle\\
        &=\frac{v_f^2}{4} \lim_{k\to 0} \langle \delta \psi_0(  \vec{k})\delta\psi_0^\dagger( - \vec{k})\rangle\\
        & = \lim_{k\to 0}\frac{v_f^2}{4} V
    \end{aligned}
    \label{eqn:type_B_QQ}
\end{equation}
Similarly $\langle \Delta Q_2 \Delta Q_2\rangle = \frac{v_f^2 V}{4} $, $\langle \Delta Q_1 \Delta Q_2\rangle = - \langle \Delta Q_2 \Delta Q_1\rangle  = \frac{iv_f^2V}{4}$.
And we also want to write down the EFT of the NGB for comparison, with $\delta \psi_0 = \frac{v_f}{\sqrt{2}}(i\pi^{(1)}-\pi^{(2)})$, and omit the $\delta \psi_{\pm} $sectors, 
\begin{equation}
    L_{0,\rm EFT} = \int d^dx \frac{v_f^2}{2}\Big[\dot \pi^{(1)} \pi^{(2)}- \pi^{(1)} \dot \pi^{(2)}\Big] -\frac{v_f^2}{4m}\Big[|\nabla\pi^{(1)}|^2+|\nabla\pi^{(2)}|^2\Big] + \frac{\chi }{2}\Big[(\dot\pi^{(1)})^2+(\dot\pi^{(2)})^2\Big]+\mathcal{O}(\pi^3)
\end{equation}
\begin{equation}
    H_{0,\rm EFT} = \int d^dx \frac{v_f^2}{4m}\Big[|\nabla\pi^{(1)}|^2+|\nabla\pi^{(2)}|^2\Big] + \frac{1}{2\chi}\Big[(\Pi_1-\frac{v_f^2\pi^{(2)}}{2})^2+(\Pi_2+\frac{v_f^2\pi^{(1)}}{2})^2\Big]+\mathcal{O}(\pi^3)
\end{equation}
where we introduce an auxiliary time-derivative term with $\chi> 0 $ to ensure that  $\langle \pi\pi\rangle $ is invertible, and take the limit $\chi\to 0$.
The canonical conjugate of $\pi^{(1,2)}$ are 
\begin{equation}
    \begin{aligned}
        \Pi_1 = \frac{v_f^2}{2}\pi^{(2)}+\chi \dot\pi^{(1)}\\
        \Pi_2 = -\frac{v_f^2}{2}\pi^{(1)}+\chi \dot\pi^{(2)}
    \end{aligned}
\end{equation}
And the Noether currents are thus
\begin{equation}
    \begin{aligned}
        &j^0_1 = \Pi_1-K_1^0=v_f^2\pi^{(2)}+\chi \dot\pi^{(1)}\\
        &j^0_2 = \Pi_2-K_2^0= -v_f^2\pi^{(1)}+\chi \dot\pi^{(2)}
    \end{aligned}
\end{equation}
which agree with the UV charges in eq.~\ref{eqn:type_B_charges}, confirming that the Goldstone coordinates are correctly identified. Here $K_{1}^0 = -\frac{v_f^2}{2}\pi^{(2)}$ and $K_{2}^0 = \frac{v_f^2}{2}\pi^{(1)}$ as in eq.~\ref{eqn:noether_current}.
With canonical commutation relations  $[\pi^a(\vec{k}),\pi^b(\vec{k'})] = 0$ and $[\pi^a(\vec{k}), \Pi_b(\vec{k'})] = i\delta^a_b (2\pi)^d\delta^{(d)}(\vec{k}+\vec{k'})$, we can solve the Heisenberg equations,
\begin{equation}
    \frac{d}{dt}
    \begin{bmatrix}
        \pi^{(1)}\\
        \pi^{(2)}\\
        \Pi_1\\
        \Pi_2
    \end{bmatrix}
    = 
    \begin{bmatrix}
         0 & \frac{-v_f^2}{2\chi} & \frac{1}{\chi} & 0 \\
         \frac{v_f^2}{2\chi} & 0 & 0 & \frac{1}{\chi}\\
         -\frac{D_k^2}{\chi} & 0 & 0 & -\frac{v_f^2}{2\chi}\\
          0 & -\frac{D_k^2}{\chi} & \frac{v_f^2}{2\chi} & 0
    \end{bmatrix}
    \begin{bmatrix}
        \pi^{(1)}\\
        \pi^{(2)}\\
        \Pi_1\\
        \Pi_2
    \end{bmatrix}
\end{equation}
where $D_k\equiv\sqrt{\frac{v_f^4}{4}+\frac{v_f^2 k^2\chi }{2m}}$.
The eigenvectors are $u_{\pm}\equiv (1, \pm i, -i D_k, \pm D_k)$, and the corresponding eigenvalues are denoted as $-i\omega_k^\pm\equiv -i\frac{D_k\pm v_f^2/2}{\chi}$. 
The field operators can be expressed as 
\begin{equation}\label{eqn:type-B_NGB_conjugates}
   \begin{bmatrix}
        \pi^{(1)}(\vec{k},t)\\
        \pi^{(2)}(\vec{k},t)\\
        \Pi_1(\vec{k},t)\\
        \Pi_2(\vec{k},t)
    \end{bmatrix} 
    =\frac{1}{2\sqrt{D_k}}\sum_{\alpha = \pm}(u_{\alpha} b_{ k}^\alpha e^{-i\omega_k^\alpha t}+u^*_{\alpha} (b_{ k}^\dagger)^\alpha e^{i\omega_k^\alpha t})
\end{equation}
Thus,
\begin{equation}
    \begin{bmatrix}
        \langle  \pi^{(1)}(\vec{k})\pi^{(1)}(\vec{k'})\rangle &  \langle  \pi^{(1)}(\vec{k})\pi^{(2)}(\vec{k'})\rangle\\
        \langle  \pi^{(2)}(\vec{k})\pi^{(1)}(\vec{k'})\rangle &  \langle  \pi^{(2)}(\vec{k})\pi^{(2)}(\vec{k'})\rangle
    \end{bmatrix}
    =
    \begin{bmatrix}
       \frac{1}{2D_k} &  0\\
       0 &  \frac{1}{2D_k}
    \end{bmatrix}
      (2\pi)^d\delta^{(d)}(\vec{k}+\vec{k'})
      \label{eqn:type_B_pi_correlators}
\end{equation}
\begin{equation}
    \begin{bmatrix}
        \langle  j_1(\vec{k})\pi^{(1)}(\vec{k'})\rangle &  \langle  j_1(\vec{k})\pi^{(2)}(\vec{k'})\rangle\\
        \langle  j_2(\vec{k})\pi^{(1)}(\vec{k'})\rangle &  \langle  j_2(\vec{k})\pi^{(2)}(\vec{k'})\rangle
    \end{bmatrix}
    = \begin{bmatrix}
        \langle  \pi^{(1)}(\vec{k})j_1(\vec{k'})\rangle &  \langle  \pi^{(1)}(\vec{k})j_2(\vec{k'})\rangle\\
        \langle  \pi^{(2)}(\vec{k})j_1(\vec{k'})\rangle &  \langle  \pi^{(2)}(\vec{k})j_2(\vec{k'})\rangle
    \end{bmatrix}^\dagger
    =
    \begin{bmatrix}
       \frac{-i}{2}  &  \frac{v_f^2}{4D_k}\\
       -\frac{v_f^2}{4D_k} &   \frac{-i}{2}
    \end{bmatrix}(2\pi)^d\delta^{(d)}(\vec{k}+\vec{k'})
    \label{eqn:type_B_pi_j_correlators}
\end{equation}
When we take $\chi \to 0$, 
\begin{equation}
    \langle  j^0_a(\vec{k})\pi^c(-\vec{k})\rangle \langle \pi(\vec{k}) \pi(-\vec{k})\rangle^{-1}_{cd} \langle \pi^d(\vec{k}) j^0_b(-\vec{k})\rangle
    = 
    \frac{v_f^2V}{2}
    \begin{bmatrix}
        1 &i\\
        -i & 1
    \end{bmatrix}
\end{equation}
which indeed verifies eq.~\ref{eq:jtopi_general}.

\subsection{Mixed States}
For mixed states, the quantum information geometry is not unique. The SLD, Wigner-Yanase, and more general Petz metrics~\cite{Petz:1999xrh} define inequivalent geometries. On the other hand, for a fixed theory and state, the NGB two-point function determines a definite metric. The question is therefore which, if any, mixed-state quantum information geometry reproduces its inverse. In the following, we use purification to identify when the inverse relation continues to hold, and show explicitly that the SLD realizes it for type-A NGBs in a thermal state.

\subsubsection{Mixed state and thermal state}
The previous derivation of the relation  $\langle jj\rangle = \langle j\pi\rangle \langle \pi\pi\rangle^{-1}\langle \pi j \rangle$ was done within pure state cases, where eq.~\ref{eqn:pure_state_correlators} relies on the vacuum annihilation condition $a_k|0\rangle = 0$. For a general mixed state $\rho$, this simplification is lost.
To see this, let $\pi = A a+Ba^\dagger$ and an operator $O = Ca+Da^\dagger$,  and denote $\langle a^2\rangle \equiv m $ and $\langle a^\dagger a\rangle\equiv \bar n$, and then $\langle a^\dagger {}^2\rangle = m^*$, $\langle a a^\dagger \rangle\equiv \bar n+1$. Their 2-point correlators satisfy,  
\begin{equation}\label{eq:inverse_relation_condition}
   \begin{aligned}
       &\langle \O \O\rangle \langle \pi\pi\rangle - \langle \O \pi\rangle \langle \pi \O \rangle \\
       & =\text{det}\Big(
       \begin{bmatrix}
           \langle\pi \pi\rangle & \langle\pi \O\rangle \\
           \langle \O \pi \rangle & \langle \O\O\rangle
       \end{bmatrix}
       \Big) \\
       & =\text{det}\Big(
       \begin{bmatrix}
           A & B \\
           C & D
       \end{bmatrix}
        \begin{bmatrix}
           m &  \bar{n}+1\\
           \bar{n} & m^*
       \end{bmatrix}
       \begin{bmatrix}
           A & B \\
           C & D
       \end{bmatrix}^T
       \Big) \\
       &= (AD-BC)^2(|m|^2-\bar n(\bar n +1))
   \end{aligned}
\end{equation}
The inverse relation thus requires either $(|m|^2-\bar n(\bar n +1)) = 0$ or $AD-BC = 0$. 
For pure states, $m = m^* = \bar{n} = 0$, but for the general mixed state, they can be non-zero. 
In a thermal state, type-A NGB is an example of $AD-BC = 0$, with the operator $\O$ being the SLD $L\propto \pi$.
However, the inverse relation does not generally hold for type-B NGBs in a thermal state. To provide the counter example, we use the same set-up as in eq.~\ref{eqn:type-B_NGB_conjugates}, keeping $\chi>0 $ finite. For the thermal state density matrix $\rho = \frac{e^{-\beta H}}{Z}\equiv \sum_n p_n |n\rangle\langle n|$, we have
\begin{equation}
    [\rho, a ] = \tanh(\frac{\beta \omega}{2})\{\rho,a\},\; [\rho, a^\dagger ] = -\tanh(\frac{\beta \omega}{2})\{\rho,a^\dagger\}
\end{equation}
and 
\begin{equation}
Q_1(\vec{k}) = \frac{i\chi}{2\sqrt{D_k}}\Big[\omega_k^-\Big(b_{k}^{+}e^{-i\omega_k^+ t}-(b_{k}^{\dagger})^+ e^{i\omega_k^+ t}\Big)+\omega_k^+\Big(b_{k}^{-}e^{-i\omega_k^- t}-(b_{k}^{\dagger})^- e^{i\omega_k^- t}\Big)\Big]    
\end{equation}
\begin{equation}
Q_2(\vec{k}) = \frac{\chi}{2\sqrt{D_k}}\Big[\omega_k^-\Big(b_{k}^{+}e^{-i\omega_k^+ t}+(b_{k}^{\dagger})^+ e^{i\omega_k^+ t}\Big)+\omega_k^+\Big(b_{k}^{-}e^{-i\omega_k^- t}+(b_{k}^{\dagger})^- e^{i\omega_k^- t}\Big)\Big],    
\end{equation}
the SLD equation $-i[Q,\rho] = \frac{1}{2}\{L,\rho\}$ can be easily solved.

\begin{equation}
L_1(\vec{k}) = -\frac{\chi }{\sqrt{D_k}}\Big[\omega_k^-\tanh(\frac{\beta\omega_k^+}{2})\Big(b_{k}^{+}e^{-i\omega_k^+ t}+(b_{k}^{\dagger})^+ e^{i\omega_k^+ t}\Big)+\omega_k^+\tanh(\frac{\beta\omega_k^-}{2})\Big(b_{k}^{-}e^{-i\omega_k^- t}+(b_{k}^{\dagger})^- e^{i\omega_k^- t}\Big)\Big]    
\end{equation}
\begin{equation}
L_2(\vec{k}) = \frac{i\chi}{\sqrt{D_k}}\Big[\omega_k^-\tanh(\frac{\beta\omega_k^+}{2})\Big(b_{k}^{+}e^{-i\omega_k^+ t}-(b_{k}^{\dagger})^+ e^{i\omega_k^+ t}\Big)+\omega_k^+\tanh(\frac{\beta\omega_k^-}{2})\Big(b_{k}^{-}e^{-i\omega_k^- t}-(b_{k}^{\dagger})^- e^{i\omega_k^- t}\Big)\Big].    
\end{equation}
From here, we already see that $L\not\propto \pi$, and $L_{1,2}\not\in\text{span}\{\pi^1,\pi^2\}$
\begin{equation}
    \pi^1(\vec{k}) = \frac{1}{2\sqrt{D_k}}\Big[b_{k}^{+}e^{-i\omega_k^+ t}+(b_{k}^{\dagger})^+ e^{i\omega_k^+ t}+b_{k}^{-}e^{-i\omega_k^- t}+(b_{k}^{\dagger})^- e^{i\omega_k^- t}\Big]    
\end{equation}
\begin{equation}
    \pi^2(\vec{k}) = \frac{i}{2\sqrt{D_k}}\Big[b_{k}^{+}e^{-i\omega_k^+ t}-(b_{k}^{\dagger})^+ e^{i\omega_k^+ t}-b_{k}^{-}e^{-i\omega_k^- t}+(b_{k}^{\dagger})^- e^{i\omega_k^- t}\Big]    
\end{equation}
Define $R_a = \O_a - \sum_b (\langle \pi\pi\rangle^{-1}\langle \pi \O\rangle)_{ab}\pi^b$, since
\begin{equation}
    \langle R_a^\dagger R_b\rangle = \langle \O_a \O_b\rangle - \langle \O_a \pi^c\rangle \langle \pi\pi\rangle^{-1}_{cd}\langle \pi^d\O_b\rangle,
\end{equation}
The general condition of the inverse relation between any operator $\O$ and $\pi$ is when $\langle R_a^\dagger R_b\rangle = 0$.
And for the thermal state, $\langle R_a^\dagger R_a\rangle =\sum_np_n\Big|R_a|n\rangle\Big|^2= 0$ only when $R_a = 0$. However, if $\O_a\not\in \text{span}\{\pi^1,\pi^2,\ldots\}$, $R_a\neq 0$. In the thermal type-B case, we can not express $L_{1,2}$ in terms of $\pi^{1,2}$, hence the corresponding $R_{1,2}\neq 0 $, and the inverse relation doesn't hold. 

To generalize the inverse relation to mixed states, we can purify the state \cite{Cottrell:2018ash}. 
As an example, consider the thermal state. One can double the Hilbert space to make the expectation value $\langle \mathcal{O}\rangle \equiv \text{Tr}(\rho \mathcal{O})$ of a mixed state identical to the one of a pure state $|\text{TFD}\rangle\equiv \sum_n \sqrt{p_n}\;|n\rangle \otimes|n\rangle_{\rm aux}$ in the enlarged Hilbert space, 
\begin{equation}
    \langle \text{TFD}| \mathcal{O}\otimes I_{\rm aux}|\text{TFD}\rangle = \sum_n p_n\mathcal{O}_{nn} = \langle \mathcal{O}\rangle.
\end{equation} 
In general, we can construct the operator that annihilates the thermofield double (TFD) state from any operator $A$ in the original Hilbert space:
\begin{equation}
   \begin{aligned}
       & (A\otimes \mathbb{I}_{\rm aux}- \mathbb{I}\otimes \rho^{1/2}_{\rm aux}A_{\rm aux}^T\rho^{-1/2}_{\rm aux})|\text{TFD}\rangle\\
       &= \sum_{n,m}\sqrt{p_n}A_{mn}|m\rangle \otimes|n\rangle_{\rm aux}  -\sqrt{p_m}A_{nm}|n\rangle \otimes|m\rangle_{\rm aux}\\
        & = 0
   \end{aligned}
\end{equation}
With this construction, we can write down the annihilation operators
\begin{equation}
    d_1 \propto a\otimes \mathbb{I}_{\rm aux}-\mathbb{I}\otimes \rho^{1/2}_{\rm aux}a_{\rm aux}^T\rho^{-1/2}_{\rm aux}
\end{equation}
\begin{equation}
    d_2 \propto - \mathbb{I}\otimes a_{\rm aux}+\rho^{1/2}_{\rm aux}a^T\rho^{-1/2}_{\rm aux}\otimes \mathbb{I}_{\rm aux}
\end{equation}
We notice that in the TFD state, the $\pi$ field and the current $j^0$ no longer live in the subspace spanned by a single pair of creation and annihilation operators as 
\begin{equation}
     \begin{bmatrix}
        \tilde\pi \\
        \tilde j^0
    \end{bmatrix}
    =
    \begin{bmatrix}
    X  & X^{*}  \\
    Y  & Y^{*}
    \end{bmatrix}
     \begin{bmatrix}
       a \\
        a^\dagger
    \end{bmatrix}
\end{equation}
After purification, the relevant operators generally lie in the larger space spanned by $d_1$, $d_2$, $d_1^\dagger$, $d_2^\dagger$ at the linear order.
However, the relation still holds when both operators lie in the same single-pair subspace. For example, if $\tilde \pi = X d_* +X^* d_*^\dagger$ and operator $\mathcal{O} = Y d_* +Y^* d_*^\dagger$, where $d_*|TFD\rangle = 0$, then by definition they still live in the same space spanned by the same pair of creation and annihilation operators.
One explicit example is the SLD corresponding to a type-A NGB in a thermal state.
\begin{equation}
    \frac{1}{2}\{\rho, L\} = i[\rho, Q]
\end{equation}
since in a thermal state 
\begin{equation}
    [\rho, a] = (1-e^{-\beta \omega})\rho a 
\end{equation}
\begin{equation}
    [\rho, a^\dagger] = (1-e^{\beta \omega})\rho a^\dagger 
\end{equation}
\begin{equation}
    \{\rho, a\} = (1+e^{-\beta \omega})\rho a 
\end{equation}
\begin{equation}
    \{\rho, a^\dagger\} = (1+e^{\beta \omega})\rho a^\dagger 
\end{equation}
If we write $Q(\vec{k}) = -i\Big(Xa_k-X^*a_k^\dagger\Big)$, then $L(\vec{k}) = 2\tanh(\frac{\beta \omega}{2})\Big[Xa_k+X^*a_k^\dagger\Big]$. For type-A NGB fields, we already see that $L\propto \pi$.
Now,
\begin{equation}
     \begin{aligned}
         \langle     L(\vec{k})L^T(-\vec{k})\rangle = \langle   L(\vec{k})   \pi^T(-\vec{k})\rangle \langle   \pi(\vec{k})   \pi^T(-\vec{k})\rangle^{-1} \langle     \pi(\vec{k}) L^T(-\vec{k})\rangle
     \end{aligned}
     \label{eqn:inverse_SLD_is_NGB}
\end{equation}
And $\langle L(\vec{k})\pi^T(-\vec{k})\rangle = \frac{1}{2}\langle \{L(\vec{k}),\pi^T(-\vec{k})\}\rangle+\frac{1}{2}\langle [L(\vec{k}),\pi^T(-\vec{k})]\rangle$
The second term $\frac{1}{2}\langle [L(\vec{k}),\pi^T(-\vec{k})]\rangle=0$, due to the proportionality. The first term can be solved with the SLD equation
\begin{equation}
    \begin{aligned}
        &\frac{1}{2}\langle \{L_a(\vec{k}),\pi_b(-\vec{k})\}\rangle \\
        &= \frac12\text{Tr}(\{\rho,L_a(\vec{k})\}\pi_b(-\vec{k})) \\
        &= \text{Tr}(i[\rho,Q_a(\vec{k})]\pi_b(-\vec{k})) \\
        &= i\langle[Q_a(\vec{k}),\pi_b(-\vec{k})]\rangle = \lim_{k\to 0}\delta_{ab}V
    \end{aligned}
\end{equation}
so we have 
\begin{equation}
     \begin{aligned}
         \langle     L(\vec{k})L^T(-\vec{k})\rangle = \langle   \pi(\vec{k})\pi^T(-\vec{k})\rangle^{-1}   V^2
         \label{eqn:LL_pipip_relation}
     \end{aligned}
\end{equation}
Thus, for a type-A NGB in a thermal state, the SLD still realizes the inverse relation between the QFM and the Goldstone correlators. This argument does not directly extend to type-B NGBs or to general mixed states, where the SLD need not lie in the same single-pair subspace as the Goldstone field.

\subsubsection{Finite temperature and the Mermin-Wagner theorem}
At finite temperature, the pure-state relation generally no longer holds. For thermal states, the SLD provides a natural replacement. We show that 
\begin{equation}
     \langle L(\vec{k})L^T(-\vec{k})\rangle = \langle L(\vec{k})\pi^T(-\vec{k})\rangle \langle \pi(\vec{k})\pi^T(-\vec{k})\rangle^{-1}\langle \pi(\vec{k}) L^T(-\vec{k})\rangle 
\end{equation}
holds by considering the same $O(N)\to O(N-1)$ model in a thermal state,
\begin{equation}
    \begin{aligned}
        \rho = \frac{e^{-\beta H}}{Z},\;\;  Z = \text{Tr} e^{-\beta H} 
    \end{aligned}
\end{equation}
To write down the SLD explicitly, we can express the SLD equation in the eigenbasis $|m\rangle$ of the density matrix $\rho = \sum_m p_m|m\rangle\langle m|$. 
\begin{equation}
    \langle m|\frac{1}{2}\{\rho,L_a\} |n\rangle= \langle m|i[\rho,Q_a]|n\rangle = i\Big[p_m(Q_a)_{mn}  -  (Q_a)_{mn}p_n\Big] = \frac12\Big[p_m(L_a)_{mn}+(L_a)_{mn}p_n\Big]
\end{equation}
Hence,
\begin{equation}
    \begin{aligned}
        (L_a)_{mn} = \frac{2i(p_m-p_n)}{p_m+p_n}(Q_a)_{mn}
    \end{aligned}
\end{equation}
The eigenstates of the density matrix are the occupation-number states of the NGB modes,
\begin{equation}
    |m\rangle = |\{n_{a,k}\}\rangle = \prod_{a,k}\frac{\Big((a^\dagger_{k})^a\Big)^{n_{a,k}}}{\sqrt{n_{a,k}! V^{n_{a,k}}}}|0\rangle
\end{equation}
where $n_{a,k}$ denotes the occupation number of the Goldstone mode associated with $\pi^a(\vec{k})$.
In this basis, the charge and the field operator can be expressed as 
\begin{equation}
    \begin{aligned}
        &(Q_a)_{n_{a,k_{\rm IR}},n'_{a,k_{\rm IR}}} \approx -v\int d^dx \langle \{n_{a,k_{\rm IR}}\}|\Pi_a(x)| \{n'_{a,k_{\rm IR}}\}\rangle  \\
        & = iv \sqrt{\frac{V\omega_{k_{\rm IR}}}{2}}\,\Big[\sqrt{n_{a,k_{\rm IR}}}\delta_{n_{a,k_{\rm IR}},n'_{a,k_{\rm IR}}+1}- \sqrt{n'_{a,k_{\rm IR}}}\delta_{n_{a,k_{\rm IR}},n'_{a,k_{\rm IR}}-1}\Big]
    \end{aligned}
\end{equation}
\begin{equation}
    \begin{aligned}
        &(\pi^a(\vec{k}))_{n_{a,k_{\rm IR}},n'_{a,k_{\rm IR}}} =  \langle \{n_{a,k_{\rm IR}}\}|\pi^a(\vec{k})| \{n'_{a,k_{\rm IR}}\}\rangle  \\
        & = \frac{1}{v\sqrt{2V\omega_{k_{\rm IR}}}}\,\Big[\sqrt{n_{a,k_{\rm IR}}}\delta_{n_{a,k_{\rm IR}},n'_{a,k_{\rm IR}}+1}(2\pi)^d\delta^{(d)}(\vec{k}_{\rm IR} - \vec{k})+ \sqrt{n_{a,k_{\rm IR}}'}\delta_{n_{a,k_{\rm IR}},n'_{a,k_{\rm IR}}-1}(2\pi)^d\delta^{(d)}(\vec{k}_{\rm IR} + \vec{k})\Big]
        \label{eqn:thermal_pi_mn}
    \end{aligned}
\end{equation}
For this oscillator with label $a$, 
\begin{equation}
    \begin{aligned}
        Z_{a,k_{\rm IR}} = \sum_{n=0} e^{-\beta  \omega_{k_{\rm IR}}n} = \frac{1}{1- e^{-\beta \omega_{k_{\rm IR}}}}
    \end{aligned}
\end{equation}
Therefore, the thermal probability of the state $|n_{a,k_{\rm IR}}\rangle$ is 
\begin{equation}
    \begin{aligned}
        p_{n_{a,k_{\rm IR}}} = e^{-\beta \omega_{k_{\rm IR}}n_{a,k_{\rm IR}}}(1-e^{-\beta\omega_{k_{\rm IR}}})
    \end{aligned}
\end{equation}
The SLD is thus,
\begin{equation}
    \begin{aligned}
        (L_a(\vec{k}_{\rm IR}))_{n,n'} = 2v\sqrt{\frac{V\omega_{k_{\rm IR}}}{2}}\tanh(\frac{\beta \omega_{k_{\rm IR}}}{2})\Big[\sqrt{n}\delta_{n,n'+1}+\sqrt{n'}\delta_{n,n'-1}\Big]
        \label{eqn:thermal_L_mn}
    \end{aligned}
\end{equation}
And the QFM is 
\begin{equation}
    F_{ab} = \frac12\text{Tr}(\rho \{L_a,L_b\}) = v^2\omega_{k_{\rm IR}}\tanh(\frac{\beta \omega_{k_{\rm IR}}}{2})\delta_{ab}V
\end{equation}
where $L_a \equiv \frac{L_a(\vec{k}_{\rm IR})+L_a(-\vec{k}_{\rm IR})}{2}$ following the Hermitian convention.
To compare it with $\langle \pi\pi\rangle$, we still need to evaluate $\langle L\pi\rangle$ and $\langle \pi L \rangle$.
\begin{equation}
    \begin{aligned}
        &\langle L_a  \pi^b \rangle = \frac12\text{Tr}(\rho [L_a,\pi^b])+ \frac12\text{Tr}(\rho \{L_a,\pi^b\})
    \end{aligned}
\end{equation}
From eq.~\ref{eqn:thermal_pi_mn} and eq.~\ref{eqn:thermal_L_mn} , we know that $[L_a,\pi^b]=0$. And using the SLD equation, the second term can be simplified
\begin{equation}
    \frac12\text{Tr}(\rho \{L_a(\vec{k}_{\rm IR}),\pi^b(-\vec{k}_{\rm IR})\}) = \frac12\text{Tr}(\{\rho, L_a\}\pi^b) = i\text{Tr}([\rho, Q_a]\pi^b) =i\text{Tr}(\rho [Q_a,\pi^b]) = \delta_a^b V
\end{equation}
So $\langle L_a(\vec{k}_{\rm IR}) \pi^b(-\vec{k}_{\rm IR})\rangle = \delta_a^bV$.
Finally, we should evaluate $\langle \pi^a\pi^b\rangle $, so that we can check the principal relation $\langle L_a(\vec{k}_{\rm IR})L_b(-\vec{k}_{\rm IR})\rangle = \langle \pi(\vec{k}_{\rm IR}) \pi(-\vec{k}_{\rm IR})\rangle ^{-1}_{ab}V^2$.
\begin{equation}
    \begin{aligned}
        \langle \pi^a(\vec{k}) \pi^b(\vec{k'})\rangle = \sum_{n}p_n \frac{2n+1}{2v^2 \omega_k}\delta^{ab}(2\pi)^d\delta^{(d)}(\vec{k}+\vec{k'}) = \frac{\coth(\frac{\beta\omega_k}{2})}{2v^2 \omega_k}\delta^{ab}(2\pi)^d\delta^{(d)}(\vec{k}+\vec{k'})
    \end{aligned}
\end{equation} 
So indeed, 
\begin{equation}
 \langle L_a(\vec{k}_{\rm IR})L_b(-\vec{k}_{\rm IR})\rangle = \langle \pi(\vec{k}_{\rm IR}) \pi(-\vec{k}_{\rm IR}) \rangle ^{-1}_{ab}V^2 =2v^2\omega_{k_{\rm IR}}\tanh(\frac{\beta\omega_{k_{\rm IR}}}{2})\delta_{ab}V
\end{equation}
The Mermin-Wagner~\cite{Mermin:1966fe,Hohenberg:1967zz,Coleman:1973ci} theorem uses the IR singularity of NGBs correlators to establish that SSB is unstable at finite temperature in spatial dimension $d\leq 2$.  
\begin{equation}
    \langle\pi^a(\vec{x})\pi^b(\vec{x})\rangle = \int \frac{d^dk}{(2\pi)^d}\frac{\coth(\frac{\beta\omega_k}{2})}{2v^2 \omega_k}\delta^{ab}\underset{k\to0}{\sim} \int \frac{d^dk}{(2\pi)^d}\frac{\delta^{ab}}{v^2\beta k^2}
\end{equation}
As in our verification of the Coleman theorem in the pure-state type-A example, we find that $F_{ab}\propto \lim_{k_{\rm IR}\to 0} k_{\rm IR}^{2-d}$. Therefore, the quantum Fisher information grows with system size only for $d>2$, consistent with the Mermin-Wagner theorem.

\subsubsection{Strong-to-weak Spontaneous Symmetry Breaking}
When we say that a symmetry is spontaneously broken, we refer to a property of the system's phase~\cite{Wang:2026ndj}. Recently, the discussion of preserved or broken symmetry has been extended to phases characterized by different notions of symmetry. More precisely, strong-to-weak spontaneous symmetry breaking (SW-SSB) refers to the transition from a phase with strong symmetry to a phase with only weak symmetry.
The definition of strong symmetry is 
\begin{equation}
    Q\rho = q \rho
\end{equation}
, i.e., all the eigenstates appearing in the density matrix  $\rho$ must also be the eigenstates of the charge $Q$ with the same eigenvalue $q$.
And the definition of weak symmetry is 
\begin{equation}
    [Q,\rho] = 0
\end{equation}
For pure states, a weak symmetry is automatically a strong symmetry. For general mixed states, a symmetric state is weakly symmetric by definition, but not necessarily strongly symmetric.
Upon purification, SW-SSB can be understood as spontaneous symmetry breaking in the enlarged Hilbert space, with an associated NGB even though the mixed state expression is still weakly symmetric.
With weak symmetry, we can pick the basis $\{|m\rangle\}$ that diagonalizes $Q$ and $\rho$ simultaneously. The density matrix $\rho =  \sum_m p_m |m\rangle \langle m|$ can be purified in an enlarged Hilbert space with a pure state expression $|\psi\rangle = \sum_m \sqrt{p_m} |m\rangle \otimes |m\rangle_{\rm aux}$.
Define $Q_L\equiv Q\otimes \mathbb{I}_{\rm aux}$ and $Q_R\equiv\mathbb{I}\otimes Q_{\rm aux}$. 
Since weak symmetry is equivalently defined as the existence of the eigenbasis $|m\rangle$ of both $Q$ and $\rho$, we know that $Q_L\;|m\rangle\otimes|m\rangle_{\rm aux} = q_m\;|m\rangle\otimes|m\rangle_{\rm aux} = Q_R\; |m\rangle\otimes|m\rangle_{\rm aux}$.
So
\begin{equation}
    \Big(Q_L-Q_R\Big)|\psi\rangle = 0
\end{equation}
$Q_L-Q_R$ is an unbroken charge in the doubled Hilbert space.
Strong symmetry $Q\rho = q\rho$ is equivalent to $q_m=q, \;\forall |m\rangle$ with $p_m\neq0$, so $Q_L|\psi\rangle = q|\psi\rangle=Q_R|\psi\rangle$. In other words, $Q_L+Q_R$ is also unbroken. 
SW-SSB is a process of breaking the symmetry corresponding to $Q_L+Q_R$. And when the weak symmetry is also broken, $Q_L-Q_R$ is broken.
Returning to the breakdown of the inverse relation between the Zamolodchikov metric and the QFM, the additional NGB associated with SW-SSB may restore the completeness of the Goldstone sector required for the inverse relation in the purified Hilbert space.

\section{Relation to the Quantum Central Limit Theorem}\label{app:QCLT}

In the context of classical measurements, the Cram\'er-Rao bound is particularly important for two reasons: (1) it is saturated by a Gaussian distribution and (2) the central limit theorem ensures most systems will nearly saturate the bounds. Naturally, this raises the question of whether the quantum Cram\'er-Rao bound is typically saturated by quantum measurement. 

Suppose we have a density matrix $\rho$ and we want to measure some properties about this state $\theta_a$, where $a = 1,.., n_{\theta}$. To be compatible with the quantum Fisher matrix, we assume we can measure these parameters by deforming the system with 
\beq
e^{iHt} \to e^{iHt + i\sum_a \theta_a Q_a} \ .
\eeq
Now, rather than using the quantum Fisher formalism, we want to simply prepare $N$ copies of the density matrix $\rho$ and measure a statistical average as follows: let $i = 1,..,N$ label the $N$ identical density matrices $\rho_i = \rho$ for all $i$. Now we define a $n_\theta$-dimensional pointer state $\X$, which is initially in the state $\X  = 0$, namely $|\X=0\rangle$. Now we deform the theory so that $\X$ calculates the mean of $\theta_a$, namely
\beq
\lim_{N \to \infty} \frac{1}{N} X_a = \theta_a \ .
\eeq
To do this, we write the reduced density matrix for the pointers as 
\beq
\rho(\X,\X') = {\rm Tr}_{\rho^{\otimes N}} e^{i \sum_{a = 1}^d \sum_{i=1}^N  P_a  Q_{a,i}} |\X= 0\rangle \langle \X' = 0 | \times \rho^{\otimes N}  e^{-i \sum_{i = 1}^d \sum_{j=1}^N P_a Q_{a,i}}
\eeq
So far this description has been general. However, we now want to define $\rho$ in a diagonal basis of $Q_a$. This is possible if $[Q_a, Q_b]  = 0$. 

Now we simply take the trace in the $Q_{i,j}$-basis so that $\rho_j \to \rho(Q_{a,i},Q'_{a',i'})$. Similarly, we can work in the momentum basis for $X$ and Fourier transform back. The result is that 
\beq
\rho(X,X') = \int \frac{d^d P d^d P'}{(2\pi)^{2d}} e^{-i \P\cdot \X + i \P' \cdot \X' } \left({\rm Tr}[ e^{i \vec P \cdot \vec Q} \rho e^{-i P' \cdot \vec Q}]\right)^N
\eeq
Now because everything is diagonal, in the $Q_{a,i}$ basis, we can replace $\rho(\vec Q,\vec Q) \to P(Q)$ as a classical probability distribution.

The actual statement is that the QCR bound can be saturated when ${\rm Tr}\rho [L_a,L_b] =0$. This is slightly more difficult to prove because $[L_a,L_b] \neq 0$ prevents us from diagonalizing the density matrix, even when $U_{ab} =0$. To see this, we again note that $\hat \theta_a = F^{-1}_{ab}  L_b$ is an unbiased estimator, as 
\beq
{\rm Tr} \rho \hat \theta_a = F^{-1}_{ab} {\rm Tr} \partial_{\theta_b} \rho = 0
\eeq
and 
\beq
\frac{1}{2} {\rm Tr}\rho \{ L_a,  \hat \theta_b\} = \frac12  F^{-1}_{bc} {\rm Tr} \rho \{L_a, L_c\} = F^{-1} F = \delta_{ab} \ .
\eeq
We will now define our walk in terms of $\hat \theta_a$ so that
\beq
\rho(X,X') = \int \frac{d^d P d^d P'}{(2\pi)^{2d}} e^{-i \P\cdot \X + i \P' \cdot \X' } \left({\rm Tr}[ e^{i \vec P \cdot \hat \theta} \rho e^{-i P' \cdot \hat \theta}]\right)^N
\eeq
Now notice that for $N$ copies, we are calculating 
\beq
\left({\rm Tr} e^{i \sum_a P_a \hat \theta_a}\rho e^{-i \sum_a P'_a \hat \theta_a}\right
)^N \ .
\eeq
Expanding the expectation value in $\vec P,\vec P'$, we get
\bea
{\rm Tr} e^{i \sum_a P_a \hat \theta_a}\rho e^{-i \sum_a P'_a \hat \theta_a}
&=& 1 -\frac{1}{2} (P-P')_a F^{-1} (P-P')_b  - i P_a (F^{-1} U F^{-1})_{ab} P'_b +{\cal O}(P^{3}) 
\eea
Notice that the $P_a P_b U^{ab} =P'_a P'_b U^{ab}  = 0$ because $U_{ab}$ is anti-symmetric by construction. This same result can be derived using the Baker–Campbell–Hausdorff formula, $e^X e^Y = \exp(X+Y +\frac{1}{2}[X,Y] +\ldots)$. 

Now, anticipating $(\X-N\vec \theta)^2 \propto N$, we expect $\vec P, \vec P' ={\cal O}(N^{-1/2})$, so taking the $N$-th power and then $N\to \infty$
\beq
\lim_{N\to \infty} \left({\rm Tr} e^{i \sum_a P_a \hat \theta_a}\rho e^{-i\sum_a P'_a \hat \theta_a}\right)^N = \exp\left(-\frac{1}{2} N[(P-P')_a F^{-1} (P-P')_b  + 2i P_a (F^{-1} U F^{-1} )_{ab} P_b' ] \right) \ .
\eeq
When $U_{ab} = 0$, this is just a real Gaussian and performing the Fourier transform gives us
\beq
\rho(X,X') = \int d^dP_+ d^d P_- e^{-\frac{i}{2}[ \vec P_+\cdot (X-X')+ \vec P_-\cdot (X+X')] } e^{-\frac{N}{2} \vec P_-^T F^{-1} P_-} = \delta(\vec X-\vec X') \exp\left(- \frac{1}{2N} \vec X^T F \vec X \right) \ ,
\eeq
where we defined $\vec P_\pm = \vec P \pm \vec P'$. We see that when $U=0$, we recover a diagonal density matrix with covariance $N(F^{-1}_{ab})$ where $F_{ab} = \frac{1}{2}{\rm Tr} \rho \{L_a,L_b\}$, thereby the estimator $X/N$ saturates the Cram\'er-Rao bound.

When $U_{ab}\neq 0$, we can see that the Fourier transform is no longer simply a function of $\vec P -\vec P'$ and therefore contains off-diagonal terms in $\rho(X,X')$. However, when $U_{ab} \neq 0$, the variables don't commute and therefore we can't work in a state for $\vec X$ without causing the variance to diverge for one of the observables. Therefore, we will assume an initial Gaussian state for $X$, $\psi(X) \propto \exp(-X^2/(4\sigma^2))$ and therefore in momentum space we have
\beq
\rho_0(\vec P,\vec P') = (8\pi \sigma^2)^{d/2} \exp\left(- \sigma^2(\vec P^2 +\vec P'{}^2)\right) = (8\pi \sigma^2)^{d/2} \exp\left(- \frac{\sigma^2}{2}(\vec P_-^2 +\vec P_+{}^2)\right)
\eeq
At this stage, we can make our life a bit easier to notice that the covariance for $\X$ when $\X-\X'=0$ is determined by 
\begin{equation}
{\cal{K}}(P_-) = \log\left(\langle e^{i \vec P_-\cdot \X} \rangle_{\vec X=\vec X'}\right) \qquad \Sigma_{ab}= - \frac{\partial^2 }{\partial P_-^a \partial P_-^b}{\cal{K}}(\vec P_-).
\end{equation}

Therefore, to determine the covariance, we simply need to integrate out $\vec P_+$,
\beq
\langle e^{i \vec P_-\cdot \X} \rangle_{\vec X=\vec X'} = C \int \frac{d^d P_+}{(2\pi)^d} \exp\left(-\frac{1}{2} N(P_- F^{-1} P_-)  - \frac{i}{2} P_- N (F^{-1} U F^{-1} )_{ab} P_+ - \frac{\sigma^2}{2}(\vec P_-^2 +\vec P_+{}^2 )\right)
\eeq
Completing the square and integrating out $\vec P_+$ we see that 
\beq
\Sigma_{ab} = N(F^{-1})_{ab} + \sigma^2 \delta_{ab} + \frac{N^2}{4\sigma^2} ( {\cal W}^T {\cal W} )_{ab} \qquad {\cal W} \equiv F^{-1} U F^{-1} \ .
\eeq
First of all, we see that the fact the observables don't commute make it impossible to set $\sigma^2 = 0$, as we did when $U_{ab} = 0$. Instead, we can minimize ${\rm Tr} \Sigma$ with respect to $\sigma^2$ to find
\beq
\Sigma_{ab,{\rm min}} = N (F^{-1})_{ab} + \frac12 N ({\rm Tr } ({\cal W}^T {\cal W})/n_{\theta})^{1/2}  \delta_{ab} + \frac{N}{2({\rm Tr }( {\cal W}^T {\cal W})/n_{\theta})^{1/2}}   ( {\cal W}^T {\cal W} )_{ab}\ .
\eeq
For a single type-B NGB pair ($n_{\theta}=2$) with identity weight, 
\begin{equation}
    \frac{\text{Tr}\Sigma_{\rm min}}{N} = \Big[\text{Tr}F^{-1}+ \sqrt{d\,{\rm Tr } ({\cal W}^T {\cal W})} \Big] = \text{Tr}F^{-1}+||F^{-1}UF^{-1}||_1,
\end{equation}
so this specific construction does saturate the Holevo-Cram\'er-Rao bound. In general, this does not saturate the Holevo-Cram\'er-Rao bound, but we see the same structure appearing. Clearly, we could work to further optimize the initial state to incrementally lower the bound (for example, allowing $\sigma^2$ to be different for each component of $\vec X$). However, as an illustration of both how the Cram\'er-Rao bound is saturated when $U_{ab} = 0$ and why it is weakened for $U_{ab} \neq 0$, the central limit theorem provides a clear demonstration of these important concepts.

\end{document}